\documentclass[a4paper,twocolumn,unpublished,11pt]{quantumarticle}
\pdfoutput=1
\usepackage[utf8]{inputenc}
\usepackage[english]{babel}
\usepackage[T1]{fontenc}
\usepackage{amsmath}
\usepackage{amssymb}

\usepackage[dvipsnames]{xcolor}
\usepackage[urlcolor=black, linkcolor=NavyBlue,citecolor=PineGreen]{hyperref}
\usepackage[numbers,sort&compress]{natbib}

\usepackage{tikz}
\usepackage{lipsum}

\newcommand\ket[1]{\left|#1\right\rangle}
\newcommand\bra[1]{\left\langle #1 \right|}

\begin{document}

\title{Exponential precision by reaching a quantum critical point}

\author{Louis Garbe}
\affiliation{Vienna Center for Quantum Science and Technology, Atominstitut, TU Wien, 1040 Vienna, Austria}
\author{Obinna Abah}
\affiliation{School of Mathematics, Statistics and Physics, Newcastle University, Newcastle upon Tyne NE1 7RU, United Kingdom}
\affiliation{Centre for Theoretical Atomic, Molecular and Optical Physics, Queen's University Belfast, Belfast BT7 1NN, United Kingdom}
\author{Simone Felicetti}
\affiliation{Istituto di Fotonica e Nanotecnologie, Consiglio Nazionale delle Ricerche (IFN-CNR),\\ Via Cineto Romano 42, 00156 Rome, Italy}
\author{Ricardo Puebla}
\affiliation{Instituto de F{\'i}sica Fundamental, IFF-CSIC, Calle Serrano 113b, 28006 Madrid, Spain}
\affiliation{Centre for Theoretical Atomic, Molecular and Optical Physics, Queen's University Belfast, Belfast BT7 1NN, United Kingdom}
\email{rpueblaantunes@gmail.com}
\orcid{0000-0002-1243-0839}

\maketitle

\begin{abstract}
Quantum metrology shows that by exploiting nonclassical resources it is possible to overcome the fundamental limit of precision found for classical parameter-estimation protocols. 
The scaling of the quantum Fisher information-- which provides an upper bound to the achievable precision-- with respect to the protocol duration is then of primarily importance to assess its performances. In classical protocols the quantum Fisher information scales linearly with time, while typical quantum-enhanced strategies achieve a quadratic (Heisenberg) or even higher-order polynomial scalings.
Here we report a protocol that is capable of surpassing the polynomial scaling, and yields an exponential advantage. Such exponential advantage is achieved by approaching, but without crossing, the critical point of a quantum phase transition of a fully-connected model in the thermodynamic limit. The exponential advantage  stems from the breakdown of the adiabatic condition close to a critical point. As we demonstrate, this exponential scaling is well captured by the new bound derived in~\cite{Garbe21}, which in turn allows us to obtain approximate analytical expressions for the quantum Fisher information that agree with exact numerical simulations. In addition, we discuss the limitations to the exponential scaling when considering a finite-size system as well as its robustness against decoherence effects.  Hence, our findings unveil a novel quantum metrological protocol whose precision scaling goes beyond the paradigmatic Heisenberg limit with respect to the protocol duration.
\end{abstract}

The existence of quantum fluctuations limits the precision that can be achieved in parameter-estimation protocols when using a limited number of physical resources. While the development of quantum mechanics has unveiled this fundamental limitation, it also provides us with the solution: By exploiting quantum resources, such as squeezing or entanglement, it is possible to overcome the fundamental limit of precision found for classical protocols~\cite{Giovannetti:04,Giovannetti:06,Paris:11,Pezze:18}. Accordingly, quantum sensing is one of the most promising applications of current quantum technologies~\cite{Dowling:03}.
In the context of quantum metrology, in order to assess the performances of a parameter-estimation strategy it is of utmost relevance to understand the scaling of the precision with respect to the amount of used resources, such as number of probes or total measurement time. At theoretical level, the most common figure of merit to quantify the estimation precision is the quantum Fisher information (QFI), which sets the ultimate achievable precision according to the Cram{\'e}r-Rao bound~\cite{Braunstein:94,Paris:11}. A paradigmatic example consists in the celebrated Heisenberg limit, valid under a series of very general assumptions~\cite{Giovannetti:06}, for which the QFI scales quadratically with the number of particles of the probe system and with the protocol duration time. This corresponds to a quadratic enhancement over the classical case, where the QFI scales linearly with those resources. It has also been shown that even super-Heisenberg scalings can be achieved when allowing for $k$-body interaction terms~\cite{Boixo:07,Roy:08}, in which case scalings $N^{2k}$, or potentially even exponential scaling in $N$, could be achieved. Another possibility is to enable time-dependent Hamiltonian evolutions~\cite{Pang:17}, which allows one to surpass the quadratic scaling in time.


In this context, systems undergoing critical phase transitions are ideal candidates for sensing applications, thanks to their high sensitivity to external perturbations. In particular, quantum phase transitions (QPT)~\cite{Sachdev} represent a compelling resource~\cite{Zanardi:08,bina_dicke_2016,FernandezLorenzo:17} for quantum metrology due to highly nonclassical properties developed in proximity of the critical point. It has also been shown~\cite{Rams:18,Garbe:20} that, in spite of the critical slowing down, the framework of critical quantum metrology makes it possible to achieve the Heisenberg scaling, where the QFI grows quadratically in time and number of probes.
Recent works~\cite{Garbe:20,Ivanov:20,Chu:21,DiCandia:21,Ilias:21,hu2021, fallani2021learning,xie2021,Niezgoda2021} have shown that the framework of critical quantum metrology can be applied to a broad class of quantum optical models. Current solid-state and atomic technology allow for the implementation of these models in a controllable way, where  their parameters can be tuned in real time. 
Intense research efforts~\cite{Garbe21,Gietka21,Montenegro2021,ivanov2021enhanced,gietka2021squeezing,Gietka_inverted,Weiss21,Yamamoto:21,Cosco:21} are now dedicated to identifying the optimal control strategies that maximize the estimation precision and mitigate potential errors.



In this article, we present the design of a quantum sensing protocol that goes beyond the typical power law scaling of the QFI, achieving an exponential scaling with respect to protocol duration time. The protocol is based on a quantum-control strategy that exploits the critical nature of quantum phase transition, by bringing cyclically the probe system in proximity of the critical point. This modulation is intrinsically nonadiabatic, as at the critical point the gap closes. We provide an explanation to the exponential time-scaling of the QFI using the bound recently introduced in Ref.~\cite{Garbe21}, showing that it is made possible by the exponential growth of the number of excitations generated by the control strategy. This general understanding also covers the results of a recent work where exponential scaling of the QFI is achieved~\cite{Gietka21}. Furthermore, in order to characterize the performances of the proposed protocols for practical applications, we analyze the effect of dissipative processes and of  finite-size corrections. We find the protocol to be resilient to thermal and dissipative effects, and the exponential scaling is preserved when the decay rate is comparable to or smaller than the single-cycle time. Finite-size corrections fix the saturation limit, and so they constrain the maximum number of cycles allowed.  The proposed protocol can be advantageous when the estimation time is the most relevant resource, and when using quantum platforms where ground-state cooling is challenging.

\section{Critical fully-connected model}\label{s:sys}
Among the different families of quantum critical systems, we focus on fully-connected models such as the quantum Rabi, the Lipkin-Meshkov-Glick~\cite{Lipkin:65}, and Dicke model~\cite{Dicke:54}. In the thermodynamic limit, these systems feature a QPT that divides the phase diagram in a normal and a symmetry-broken phase~\cite{Ashhab:13,Bakemeier:12,Hwang:15,Puebla:16,Ribeiro:07,Ribeiro:08,Emary:03prl,Emary:03,Lambert:04}. In the Dicke and Lipkin-Meshkov-Glick models, the thermodynamic limit refers to the standard notion of infinitely many components. In the quantum Rabi model,
 and related finite-component critical systems~\cite{Liu:17,Shen:17,Peng:19,Zhu:20,Felicetti:20,Shen:21}, it rather refers to a certain ratio of the system parameters. For instance in the quantum Rabi model, a QPT emerges when the qubit frequency becomes much larger than the field frequency. The frequency ratio acts as an effective system size, and finite-size critical exponents can also be defined in this case.
 These fully-connected systems admit a simple description in terms of an effective bosonic mode whose potential depends on a rescaled and dimensionless coupling strength $g$. These models
  constitute a suitable test-bed for the exploration of different aspects of quantum critical phenomena~\cite{Dusuel:04,Vidal:06,Bastidas:12,Puebla:13,Brandes:13,Puebla:15,Puebla:17,Liu:17,Shen:17,Garbe:17,Wang:18,Zunkovic:18,Hwang:18,Hwang:19,Felicetti:20,Zhu:20,Puebla:20,Puebla:20b,Corps:21}. Within the normal phase $0\leq g\leq 1$ and in the thermodynamic limit, the effective Hamiltonian describing these systems can be written as~\cite{Hwang:15,Puebla:20,Garbe21}
\begin{align}\label{eq:H0}
H=\omega a^\dagger a- \frac{g^2\omega}{4}(a+a^\dagger)^2,
\end{align}
where $[a,a^\dagger]=1$ and $\omega$ denotes the frequency of the bosonic mode. Here, $g$ is a rescaled coupling parameter which can typically be written as $g=\lambda/\lambda_c$, where $\lambda$ is the dimensionful coupling strength of the model and $\lambda_c$ its critical value.
From Eq.~\eqref{eq:H0} it follows that the energy gap vanishes at the QPT as $\Delta(g)\propto |g-g_c|^{1/2}$ so that the critical exponents in this case are of a mean-field type $z\nu=1/2$~\cite{Sachdev}. It is worth stressing that the critical traits in these systems have been experimentally observed~\cite{Zibold:10,Baumann:10,Baumann:11,Mottl:12,Jurcevic:17,Cai:21}.
Importantly, the critical coupling value $\lambda_c$ typically depends on $\omega$. In the Lipkin-Meshkov-Glick and quantum Rabi model, for example, we have a dependence as $\lambda_c\propto \sqrt{\omega}$. Therefore, the rescaled coupling depends itself on $\omega$ as $g\propto 1/\sqrt{\omega}$. In the following, we assume this dependence, which makes the sensing protocol critically dependent on the value of $\omega$~\cite{Garbe21}.

\section{Quantum Fisher Information}\label{s:QFI}
The ultimate precision for the estimation of a parameter $x$ is given by the QFI~\cite{Paris:11}, denoted as $I_x$, such that the variance for the estimated parameter $x$ is bounded as $(\delta x)^2\geq I_x^{-1}$ for a single measurement, which is known as the quantum Cram{\'e}r-Rao bound.  This result is obtained optimizing over all possible positive operator-valued measurements and classical data processing. Thus, the scaling of $I_x$ with respect to the experimental resources, such as the duration of the metrological protocol, is of key importance. 
Let us denote by $\rho_x$ the system state in which the unknown value of $x$ has been encoded. The QFI is related to the Bures distance between two infinitesimally closed states, $\rho_x$ and $\rho_{x+\epsilon}$, which can be written as $d_{{\rm B},x}^2=2(1-{\rm Tr}[\sqrt{\sqrt{\rho_x}\rho_{x+\epsilon}\sqrt{\rho_x}}])$, so that~\cite{Pinel2013}
\begin{align}
I_x=4\left(\left.\frac{\partial d_{{\rm B},x}}{\partial \epsilon}\right|_{\epsilon=0} \right)^{2}.
\end{align}
The signal-to-noise ratio can be then written as $Q_x=x^2 I_x$. 

Since the Hamiltonian~\eqref{eq:H0} is quadratic in $a$ and $a^\dagger$, any initial Gaussian state evolving under~\eqref{eq:H0} is also a Gaussian state~\cite{Ferraro}. Recall that a Gaussian state $\rho$ is that whose Wigner function is Gaussian, and thus $\rho$ can be fully determined in terms of its first and second moments in the two-dimensional phase space ${\bf X}^{\top}=(x,p)$~\cite{Ferraro}, where here we employ the convention $x=a+a^\dagger$ and $p=i(a^\dagger-a)$. The first moments of the state is simply $\langle {\bf X}^\top\rangle=({\rm Tr}[\rho \ x],{\rm Tr}[\rho \ p])$. The second moments are given by the covariance matrix ${\bf R}$, which is real and symmetric, and its matrix elements read as
\begin{align}
R_{i,j}=\frac{1}{2}\langle X_i X_j+X_jX_i\rangle-\langle X_i\rangle\langle X_j\rangle.
  \end{align}

As shown in Ref.~\cite{Pinel2013}, the QFI adopts the following form for Gaussian states,
\begin{align}\label{eq:Ix}
I_x=\frac{1}{2}\frac{{\rm Tr}[({\bf R}^{-1}\partial_x {\bf R})^2]}{1+P^2}+2\frac{(\partial_x P)^2}{1-P^4}+L_x,
\end{align}
with $L_x={\bf \Delta  X}'^{\top}_x{\bf R}^{-1}{\bf \Delta X}'_x$ and ${\bf \Delta X}'_x=\partial \langle {\bf X}_{x+\epsilon}-{\bf X}_{x}  \rangle/\partial\epsilon|_{\epsilon=0}$, and $P={\rm det}[{\bf R}]^{-1/2}$ denotes the purity of the Gaussian state $\rho$. Throughout the article we will consider Gaussian states with $\langle x\rangle=\langle p\rangle=0$, and therefore $L_x=0$ in Eq.~\eqref{eq:Ix}. In this manner, the QFI and the corresponding signal-to-noise ratio $Q_x$ can be computed from the covariance matrix ${\bf R}$.  

In the following we will focus on the estimation of the bosonic frequency $\omega$, although similar results can be found for the estimation of $g$ in Eq.~\eqref{eq:H0}. 
\begin{figure}[t]
  \centering
  \includegraphics[width=\linewidth]{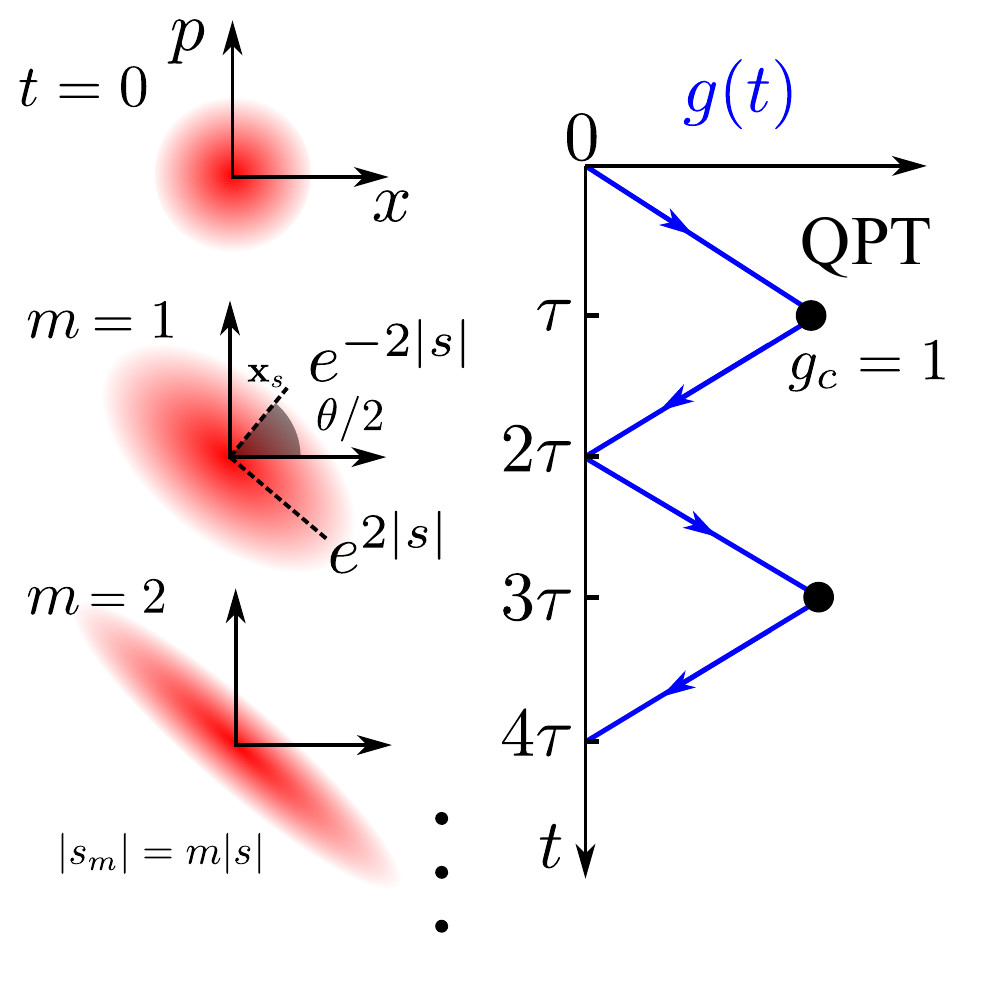}
  \caption{Schematic illustration of the protocol. At $t=0$ the state is assumed to be in a vacuum or thermal state, which is brought to the critical point $g_\tau=g_c=1$ in a time $\tau$ by tuning $g(t)$. The cycle is completed at time $2\tau$ such that $g(2\tau)=g(0)=0$. After one cycle $m=1$ the state becomes squeezed, with a squeezing parameter and angle $|s|$ and $\theta$, respectively, so that the state is squeezed along the direction ${\bf x}_s^\top=(\cos(\theta/2),\sin(\theta/2))$ in the phase space ${\bf X}^\top=(x,p)$, reducing the variance by a factor $e^{-2|s|}$. Performing another cycle, $m=2$, the state can be further squeezed such that $|s_m|=m|s|$ where $m=1,2,\ldots$, under suitable parameters. This squeezing amplification leads to an exponential precision for the estimation of the system parameters (see main text for further details). }
  \label{fig1}
\end{figure}

\section{Non-adiabatic cycles: Exponential scaling}\label{s:dyn}
The Hamiltonian in Eq.~\eqref{eq:H0} displays a QPT at the critical point $g_c=1$, which is accompanied by a vanishing energy gap, among other features. As a consequence, by tuning $g(t)$ towards the QPT in a finite time the system will unavoidably depart from adiabaticity~\cite{Polkovnikov:08,Defenu:20}. 
Such non-adiabaticity translates in the formation of quantum excitations in the system, which can be harnessed and beneficial in different contexts~\cite{Abah:21}. In the following, we show that the non-adiabaticity caused by the QPT can be exploited to lead in an exponential scaling of the QFI with respect the protocol duration.

In particular, we choose a protocol $g(t)$ that completes a cycle in a time $2\tau\gtrsim 1/\omega$ as (cf. Fig.~\ref{fig1})
\begin{align}\label{eq:gt}
  g(t)=\begin{cases} g_\tau \frac{t}{\tau} &\ {\rm for} \ \ 0\leq t\leq \tau \\
  g_\tau\left(2-\frac{t}{\tau}\right) &\ {\rm for} \ \ \tau \leq t\leq 2\tau,
  \end{cases} 
  \end{align}
 so that $g(0)=0$ and $g(\tau)=g_\tau$. The state at any time $t$ follows from $\dot{\rho}=-i[H(t),\rho(t)]$, with the initial and final state upon the completion of the cycle given by $\rho(0)$ and $\rho(2\tau)$, respectively. Note that the condition $2\tau\gtrsim 1/\omega$ rules out fast cyclic transformation ($\omega\tau\rightarrow 0$) for which the initial state remains trivially unchanged.
 If we perform a cycle away from the critical point, i.e. for $g_\tau<g_c=1$, and for sufficiently slow cycles, $\tau\gg 1/\Delta(g_\tau)$, the protocol is able to meet the adiabatic condition. Hence, by virtue of the adiabatic theorem, the state upon the cyclic transformation is simply $\rho(2\tau)=\rho(0)$. By contrast, if we bring the system all the way to the critical point, i.e. if $g_\tau=g_c=1$, the adiabatic condition will break down at some point, since $\Delta(g_c)=0$. Thus, $\rho(2\tau)\neq \rho(0)$ regardless of how slow the cycle is performed, as studied in Ref.~\cite{Defenu:20}.
 Indeed, the transformation~\eqref{eq:gt} reaching the critical point produces squeezing, that is, $\rho(2\tau)=\mathcal{S}(s)\rho(0)\mathcal{S}^\dagger(s)$ with $\mathcal{S}(s)={\rm exp}[(s (a^\dagger)^2-s^*a^2)/2]$ and $s=|s|e^{i\theta}$, with $|s|$ and $\theta$ the squeezing parameter and its angle, respectively. In this manner, the state is squeezed along ${\bf x}_s^\top=(\cos(\theta/2),\sin(\theta/2))$ in the phase space. To a good degree of approximation  (cf. App.~\ref{app:s}), the acquired squeezing after completing the protocol $g(t)$ with duration $2\tau$ (and $\tau\gtrsim 1/\omega$) is given by~\cite{Defenu:20,Abah:21} 
\begin{align}\label{eq:s}
|s|=\frac{\log(3)}{2},
\end{align}
while the angle $\theta$ depends on $\tau$ (see Fig.~\ref{fig1}). 
Note that this is caused solely by the presence of the QPT. 

Although setting $g_\tau=1$ allows to automatically break adiabaticity, the resulting squeezing is robust against small deviations from $g_\tau=1$, i.e. for cycles with $|g_\tau-g_c|\ll 1$ Eq.~\eqref{eq:s} still holds for $1/\omega\lesssim \tau \lesssim 1/\Delta(g_\tau)$ (cf. App.~\ref{app:s}).

\begin{figure}[t!]
  \centering
  \includegraphics[width=\linewidth]{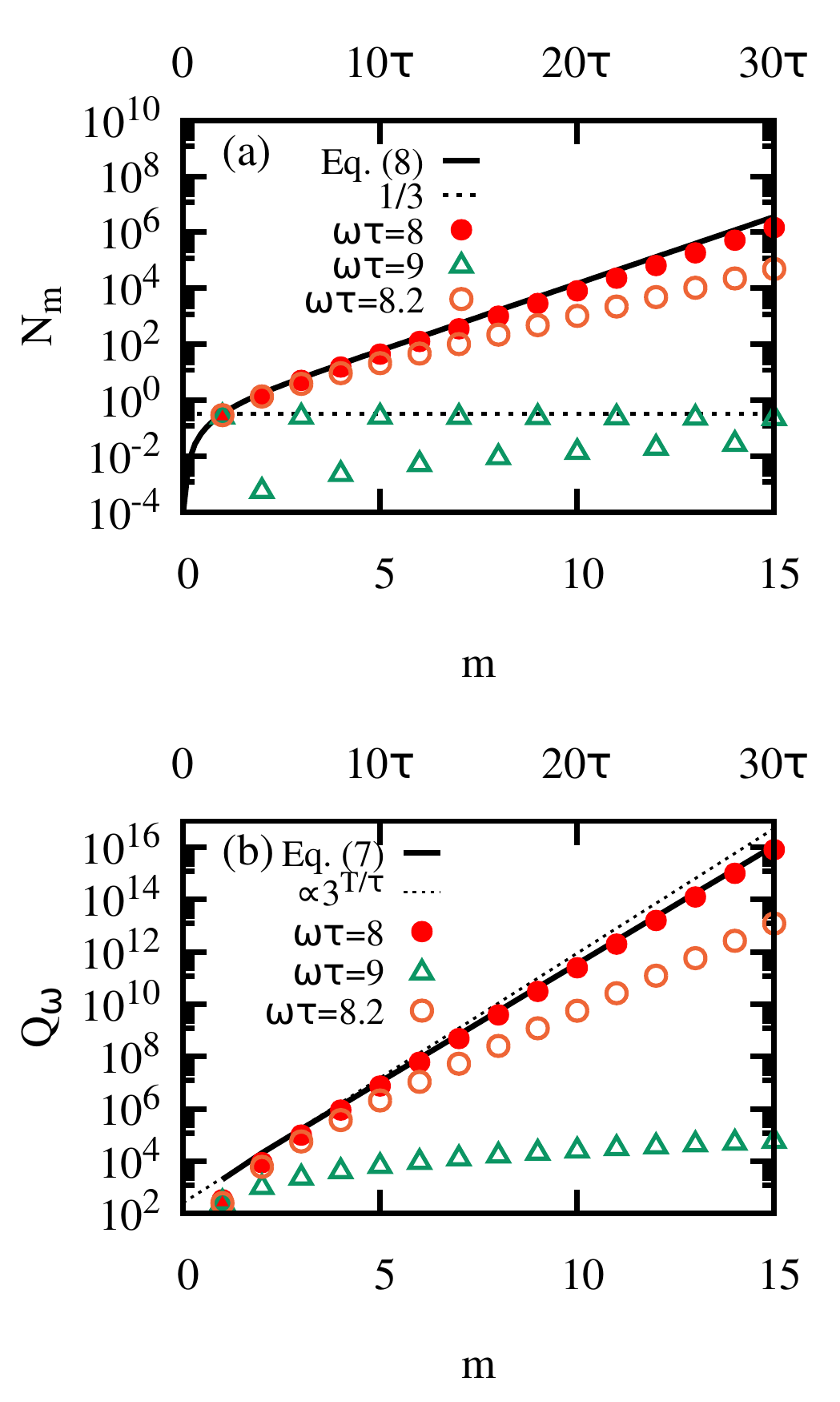}
  \caption{Noiseless dynamics in the thermodynamic limit for an initial vacuum state. Panel (a) shows the number of bosons after $m$ cycles $N_m$ for $\omega\tau=8$ (solid red circles) $\omega\tau=9$ (open green triangles) and $\omega\tau=8.2$ (open orange circles). The phase-matching condition is achieved for $\omega\tau=8$ which leads to an exponential growth of $N_m$, while for $\omega\tau=9$ it follows that $N_{2m+1}\approx 1/3$ (dotted black line) (see main text). The results for $\omega\tau=8$ follow the trend predicted in Eq.~\eqref{eq:Nm} (solid black line), which is closely followed by small deviations from the phase-matching condition, as illustrated for $\omega\tau=8.2$. Panel (b) shows the computed signal-to-noise ratio $Q_\omega=\omega^2 I_\omega$ together with the bound computed numerically using Eq.~\eqref{eq:Ib} (solid black line) as well as its approximated expression (dotted black line), given in Eq.~\eqref{eq:Ibscaling}, i.e. $I_\omega^{\rm B}\approx 4\tau^2 3^{T/\tau}$. The exponential scaling is robust against deviations from the phase-matching condition. For $\omega\tau=9$ there is no exponential scaling but rather  $Q_\omega\propto (2m\tau)^2=T^2$, which is also well captured by the bound (not explicitly shown).}
  \label{fig2}
\end{figure}

This cyclic transformation can be carried out $m$ times by a concatenation of the protocol $g(t)$ in~\eqref{eq:gt}, that is, $g(t+2m\tau)=g(t)$ with $t\in[0,2\tau]$ and $m=1,2,\ldots$, in a total time $T=2m\tau$.  By doing so, the produced squeezing can be amplified to yield $|s_m|=m|s|=m\log(3)/2$, where $|s_m|$ denotes the squeezing produced on the initial state $\rho(0)$ after the $m$th cycles, i.e. $\rho(2m\tau)=\mathcal{S}(s_m)\rho(0)\mathcal{S}(s_m)$ with $s_m=|s_m|e^{i\theta_m}$ (cf. Fig.~\ref{fig1}). Such amplification requires a phase-matching condition for subsequent cycles, i.e. $\theta_{m+1}=\theta_{m}$. However, if $\theta_{m+1}=\theta_m+\pi$ (modulo $2\pi$) the $m+1$th cycle compensates the squeezing generated in the previous cycle, and thus we expect $|s_{2m+1}|\approx |s|=\log(3)/2$ and $|s_{2m}|\approx 0$, so that $N_{2m+1}\approx N_{m=1}=1/3$ while $N_{2m}\approx 0$.

It is worth mentioning that other schemes can also achieve a linear amplification of squeezing, such as the one reported in Ref.~\cite{Cosco:21} based on a suitable periodic modulation of the oscillator frequency~\cite{Rashid:16} which differs from Eq.~\eqref{eq:gt}.


\subsection{Bound to the quantum Fisher information}

As aforementioned, Eq.\eqref{eq:Ix} allows us to compute exactly the QFI. However,  before presenting the numerical results in the next section, we show here that the main features of the QFI behavior can be predicted analytically. 
For that we rely on a recently derived bound to the QFI, which is valid for active interferometric protocols and Gaussian states. This bound is denoted as $I^{\rm B}_\omega$ such that $I_\omega\leq I_\omega^{\rm B}$, and in our case is given by (see Ref.~\cite{Garbe21} for the details of the derivation):
\begin{align}\label{eq:Ib}
I_\omega^{\rm B}=8(\chi^2+\phi^2)\left[\int_0^{T}dt (2N(t)+1) \right]^2,
  \end{align}
where $N(t)$ denotes the number of bosons at time $t$, while $\chi$ and $\phi$ are the eigenvalues of the matrix $\partial_\omega H(t)$ in the phase space ${\bf X}$, which here take a time-independent value $\chi=\phi=1/2$~\cite{Garbe21}. If the number of probes $N(t)$ is time-independent, Eq.\eqref{eq:Ib} returns the Heisenberg scaling $N^2T^2$. By contrast, if $N(t)$ is time-dependent, which is the case here, we can achieve more exotic scalings in $T$. For simplicity, let us consider an initial vacuum state, $\rho(0)=\ket{0}\bra{0}$, although we remark that the results are robust against finite-temperature initial states (see App.~\ref{app:T}). From Eq.~\eqref{eq:s} and since $N={\rm Tr}[\mathcal{S}(s)\rho(0)\mathcal{S}^\dagger(s)a^\dagger a]=\sinh^2(|s|)$, we expect the number of bosons $N_m={\rm Tr}[\rho(2m\tau) a^\dagger a]$ after $m$ cycles, assuming a phase-matching condition, to obey
\begin{align}\label{eq:Nm}
N_m=\sinh^2\left(\frac{m}{2}\log(3)\right),
  \end{align}
which grows exponentially with $m$, $N_m\sim 3^m/4$ for $m\gg 1$. Hence, we can already anticipate that $I_\omega^{\rm B}$ will show a similar exponential scaling. Indeed, by approximating $N(t)$ during $t\in [2(m-1)\tau,2m\tau]$ by $N_m$ so that $\int_{(2m-1)\tau}^{2m\tau}dt \ (2N(t)+1)\approx 2\tau(2N_m+1)$, we find
\begin{align}
I_\omega^B&\leq 16\tau^2(2N_m+1)^2\nonumber \\&\approx 64\tau^2 \sinh^4(m\log(3)/2)
  \end{align}
where in the last step we have assumed $N_m\gg 1$. In this manner, for $m\gg 1$ we find
\begin{align}\label{eq:Ibscaling}
  I_\omega^B \approx 4\tau^23^{2m}=4\tau^2 3^{T/\tau},
\end{align}
where $T=2m\tau$ is the total time of the protocol after $m$ cycles. That is, the bound to the QFI scales exponentially with $T$. This is the central result of the article. In the reminder of the article we will employ numerical simulations to corroborate the validity of the previous expression (cf. Sec.~\ref{ss:dyn}), while in Sec.~\ref{s:imp} we investigate the noise robustness of such exponential scaling, as well as the potential impact of finite-size effects.

\begin{figure}[t!]
  \centering
  \includegraphics[width=\linewidth]{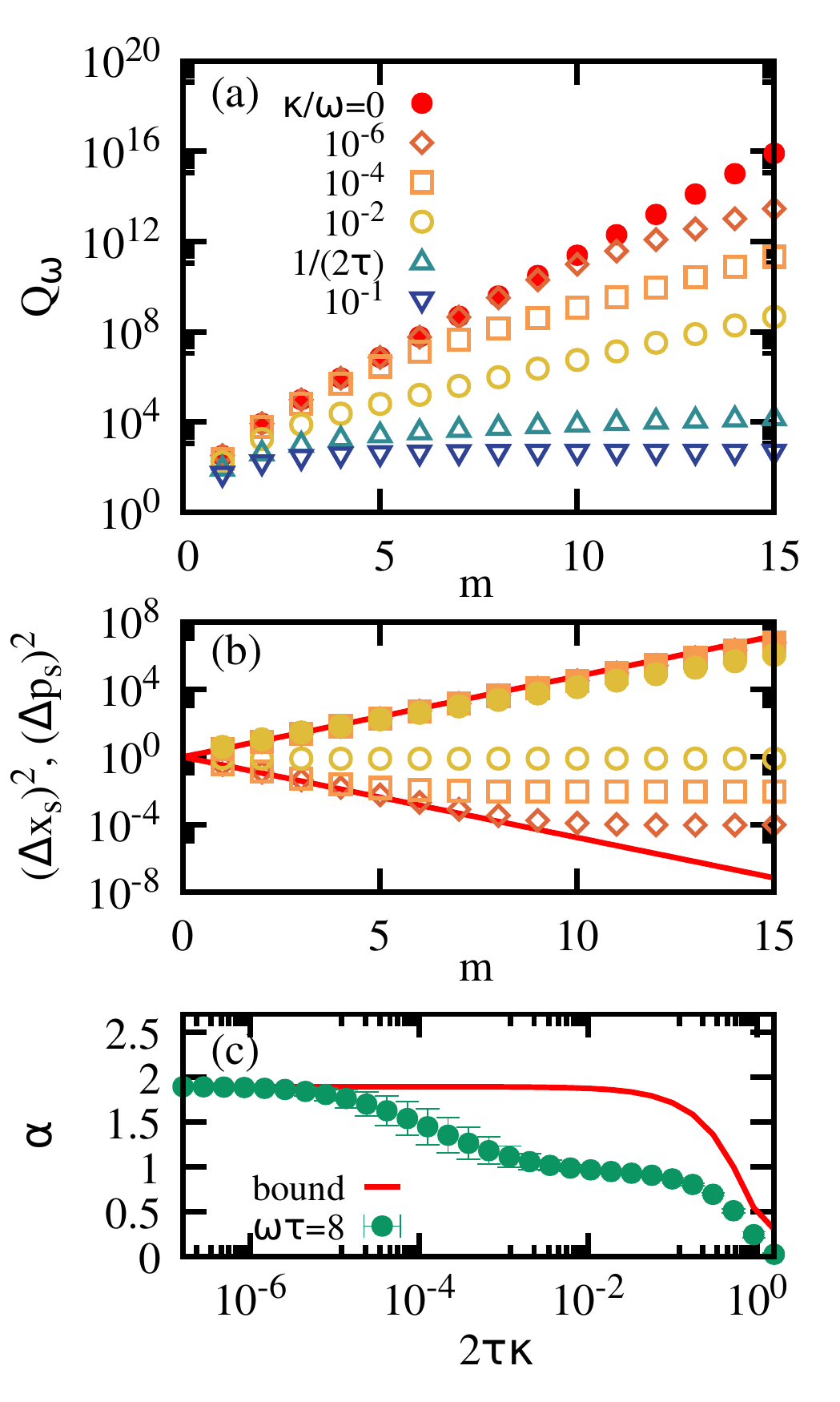}
  \caption{Robustness of the exponential scaling in $Q_\omega$ versus decoherence effects. The results have been obtained for $\rho(0)=\ket{0}\bra{0}$, $g_\tau=g_c=1$ and $\omega\tau=8$ as in Fig.~\ref{fig2}, although equivalent results can be found for different $\omega\tau\gtrsim 1$. Panel (a) shows $Q_\omega$ for increasing number of cycles $m$, (b) illustrates the variance along the squeezing axes, while (c) depicts the prefactor $\alpha$ such that $Q_\omega\propto 3^{\alpha m}$, obtained as a best fit in the interval $m\in[5,10]$, as a function of $2\tau\kappa$. In (b) the solid red line corresponds to $3^{\pm m}$, as expected theoretically for $\kappa=0$. Open (full) points correspond to $(\Delta {\bf x}_s)^2$ ($(\Delta {\bf p}_s)^2$) with the same format as in (a). In (b), the solid red line corresponds to $\alpha$ obtained from a fit to the bound in Eq.~\eqref{eq:Ib}. The exponential advantage holds even for non-zero dissipation interaction strength $\kappa$, although the prefactor $\alpha$ in the scaling decreases. For $2\tau\kappa=1$ (open yellow triangle in (a) and (b)) the scaling shifts to the standard quadratic scaling $Q_\omega\propto T^2$, while for $2\tau\kappa>1$ the signal-to-noise ratio $Q_\omega$ saturates and $\alpha=0$.}
  \label{fig3}
\end{figure}

\subsection{Noiseless dynamics in the thermodynamic limit}\label{ss:dyn}
The dynamics in the thermodynamic limit can be computed exploiting the quadratic nature of the Hamiltonian~\eqref{eq:H0}. Assuming that the initial state is such that $\langle x\rangle=\langle p\rangle=0$ and that it undergoes a noiseless evolution under the protocol $g(t)$, the state at time $t$ is completely characterized by the following time-dependent Lyapunov equation of motion for the covariance matrix (see App.~\ref{app:dyn})
\begin{align}\label{eq:Lya1}
\dot{{\bf R}}(t)={\bf W}(t) {\bf R}(t)+{\bf R}(t){\bf W}^\dagger(t),
\end{align}
with
\begin{align}
  {\bf W}(t)=\begin{bmatrix} 0 & \omega(1-g^2(t))\\ -\omega & 0
  \end{bmatrix}
  \end{align}
while $\langle x\rangle=\langle p\rangle=0\ \forall t$. The number of bosons at time $t$ is then given by $N(t)=\frac{1}{4}({\rm Tr}[{\bf R}(t)]-2)$, while the QFI can be computed exactly using Eq.~\eqref{eq:Ix}, and compared with the bound prediction \eqref{eq:Ibscaling}. In order to test the prediction of the exponential scaling for $I_\omega$ we consider $\rho(0)=\ket{0}\bra{0}$ so that ${\bf R}(0)=\mathbb{I}$. Numerically solving Eq.~\eqref{eq:Lya1} under the protocol $g(t)$ given in Eq.~\eqref{eq:gt}, we can compute $N_m=N(2m\tau)$ and the corresponding signal-to-noise ratio $Q_\omega=\omega^2 I_\omega$. The phase-matching condition is achieved here when $\omega \tau=2 n$ with $n=1,2,\ldots$, so that $\theta_m = \pm \pi/2 \ \forall m$, while for $\omega \tau=2n+1$ the $m+1$th cycle counteracts the generated squeezing in the previous one (cf. App.~\ref{app:dyn}). Thus, for the phase-matching condition the state is squeezed in the direction ${\bf x}_s^\top=(1,\pm 1)/\sqrt{2}$ in the phase space. The number of bosons after $m$ cycles is plotted in Fig.~\ref{fig2}(a) for $\omega\tau=8$ (phase-matching condition),  $\omega\tau=8.2$ and $\omega\tau=9$. The results clearly show the exponential growth of $N_m$, which is well captured by Eq.~\eqref{eq:Nm}, for $\omega\tau=8$. Finally, the signal-to-noise ratio $Q_\omega$ is plotted in Fig.~\ref{fig2}(b), together with the bound and its approximated value given in Eq.~\eqref{eq:Ib} and \eqref{eq:Ibscaling}, respectively. The numerical results show that the QFI shows an exponential scaling with the protocol duration $T$, $Q_\omega\propto 3^{T/\tau}$,  as predicted by Eq.~\eqref{eq:Ibscaling}. The exponential scaling still holds, although the prefactor is reduced, in situations when $\omega\tau\sim 2n$ but $\omega\tau\neq 2n$, as exemplified by $\omega\tau=8.2$.  To the contrary, for $\omega\tau=2n+1$ the number of bosons is bounded by $1/3$ which leads to the standard $Q_\omega\propto T^2$ scaling.   We stress again that, although the previous results have been computed considering $\rho(0)=\ket{0}\bra{0}$, i.e. a zero-temperature initial state,  initial thermal states with an arbitrary temperature also lead to an exponential advantage (cf. App.~\ref{app:T}).

\section{Noise robustness and finite-size effects}\label{s:imp}
Let us now analyze the robustness of the exponential scaling in the QFI  with respect to decoherence. For that, we model the interaction of the system with an environment at an inverse temperature $\beta$ such that $N_{\rm th}=(e^{\beta \omega }-1)^{-1}$ through the standard Lindblad master equation~\cite{Breuer}
\begin{align}\label{eq:me}
\dot{\rho}(t)=-i[H(t),\rho(t)]+\mathcal{D}[\rho(t)]
  \end{align}
where the dissipator reads as
\begin{align}
\mathcal{D}[\rho]=&\kappa\frac{(N_{\rm th}+1)}{2}\left(2a\rho a^\dagger-\{a^\dagger a,\rho\}\right)\\&+\kappa \frac{N_{\rm th}}{2}\left(2 a^\dagger \rho a-\{a a^\dagger,\rho\}\right),
\end{align}
and the parameter $\kappa$ accounts for the system-environment interaction strength. As the master equation is still quadratic in $a$ and $a^\dagger$, the time-dependent Lyapunov equation for the covariance matrix modifies to
\begin{align}\label{eq:Lya2}
\dot{{\bf R}}(t)={\tilde {\bf W}}(t) {\bf R}(t)+{\bf R}(t){\tilde {\bf W}}^\dagger(t)+{\bf F},
\end{align}
where now  ${\bf F}=\kappa(2N_{\rm th}+1)\mathbb{I}$ and ${\tilde {\bf W}}={\bf W}-\kappa/2 \mathbb{I}$. As before, we compute the $Q_\omega$ from Eq.~\eqref{eq:Ix} solving Eq.~\eqref{eq:Lya2}. The results are gathered in Fig.~\ref{fig3}(a) for $N_{\rm th}=2$, which shows that the exponential scaling is robust against decoherence provided $2\tau\kappa\ll 1$. There we show $Q_\omega$ for different values of $\kappa$ starting from $\rho(0)=\ket{0}\bra{0}$. For $2\tau\kappa\ll 1$ the dissipation still permits an exponential scaling $Q_\omega\propto 3^{\alpha m}=3^{\alpha T/(2\tau)}$ but with smaller prefactor, $0<\alpha\leq 2$, which is reduced as $\kappa$ increases. To the contrary $2\tau\kappa\gg 1$ the state relaxes to the thermal equilibrium before the cycle is completed and thus  $Q_\omega$ does not increase with the protocol duration. For $2\tau\kappa\approx 1$ both mechanisms are balanced and  $Q_\omega \propto T^2$ is still possible. Since $Q_\omega\propto 4\tau^2 3^{2m}$, for a fixed number of cycles $m$, the QFI is maximal when $2\tau\kappa=1$ provided $\omega\tau$ ensures the phase-matching condition, while for fixed total evolution time $T$ it is more beneficial to perform the largest allowed number of cycles $m=T/(2\tau)$, and hence to take the shortest possible $\tau$, with the constraints  $2\tau\gtrsim 1/\omega$ and $2\tau\kappa\ll 1$. After $m$ cycles the system finds itself in a squeezed thermal state, so that the variance along the squeezing axis reads as $(\Delta {\bf x}_s)^2=(2n_\kappa+1)e^{-2 |s_m|}$, while $(\Delta {\bf p}_s)^2=(2n_\kappa+1)e^{2 |s_m|}$ is enlarged, where $n_\kappa$ is the number of thermal bosons acquired by the state due to the interaction with the environment. Moreover, the total number of bosons after $m$ cycles is $N(2\tau m)=(2n_\kappa+1)\sinh^2|s_m|+n_\kappa$. Although $|s_m|$ increases with $m$ for $2\tau\kappa\lesssim 1$, so does the number of thermal bosons. This leads to a saturation in the variance $(\Delta {\bf x}_s)^2$, while $(\Delta {\bf p}_s)^2$ keeps increasing with $m$, which still allows $Q_\omega$ to grow. This is shown in Fig.~\ref{fig3}(b). For $2\tau\kappa>1$, both quadratures saturate, and therefore also $Q_\omega$.  In addition, we also show the behavior of the prefactor $\alpha$ as a function $2\tau\kappa$, which is illustrated in Fig.~\ref{fig3}(c). There we show the fitted $\alpha$ such that $Q_\omega\propto 3^{\alpha m}$. For a noiseless evolution, $\alpha\approx 2$ (see Fig.~\ref{fig2}(b) and App.~\ref{app:s} for a comment regarding the expected deviation with respect $\alpha=2$), while as $\kappa$ increases $\alpha\rightarrow 0$. Yet, there is a wide range of values for $\kappa$ in which the exponential scaling holds. Similar results can be found for other $\omega\tau$ fulfilling (or close to) a phase-matching condition.   For comparison, we also compute $\alpha$ using the bound in Eq.~\eqref{eq:Ib}. As shown in Fig.~\ref{fig3}(b), although the bound still captures the exponential scaling for $2\tau\kappa\lesssim 1$, it becomes loose and overestimates the prefactor $\alpha$. It is worth mentioning that the transition from $\alpha=2$ to $1$ can be understood as a departure from the Heisenberg to the standard quantum limit. Indeed, we can re-express $Q_\omega$ in terms of both the maximum number of bosons and the protocol duration time. Then, $\alpha=2$ corresponds to $Q_\omega\sim N_m^2T^2$ (Heisenberg limit), while $\alpha=1$ to $Q_\omega\sim N_m T$, which is the scaling of the standard quantum limit. This expression is useful if we are in a situation in which both the protocol duration and the number of photons are limited. By contrast, if the only relevant resource is the time, we can express again $N_m\sim 3^m=3^{T/\tau}$. Then both $\alpha=1$ and $\alpha=2$ lead to an exponential scaling of the QFI with $T$, yet $\alpha>1$ means a scaling beyond the standard quantum limit.


Finally, we turn our attention to finite-size effects. As commented in Sec.~\ref{s:sys}, the Hamiltonian $H$ in Eq.~\eqref{eq:H0} is valid in the normal phase and in the thermodynamic limit. Yet, any realistic exploration of a critical system is unavoidably accompanied by finite-size corrections. Let us denote by $\eta$ the system size, which in Lipkin-Meshkov-Glick and quantum Rabi model refers to the number of spins and a ratio between spin and bosonic frequencies, respectively. Indeed, for $\eta<\infty$ the leading-order correction to Eq.~\eqref{eq:H0} leads to a Hamiltonian of the form $H_\eta=H+\frac{f(g)}{\eta}(a+a^\dagger)^4$ where $f(g)$ is a function of $g$~\cite{Puebla:20,Garbe21}. In this manner, the $1/\eta$ correction introduces a confining potential and lifts the vanishing energy gap at $g_c=1$. In our case, such correction can become significant since $N$ grows exponentially, and so does ${\rm Tr}[\rho(t)(a+a^\dagger)^4]$. Hence, we expect that the exponential scaling reported in Sec.~\ref{s:dyn} holds as long as $f(g) {\rm Tr}[\mathcal{S}(s_m)\rho(0)\mathcal{S}^\dagger(s_m)(a+a^\dagger)^4]\ll \eta$, as otherwise the correction can no longer considered as a perturbation and will significantly modify the Gaussian nature of the state and, consequently, the generated squeezing. The previous conditions allow us to define a maximum number of cycles $m^*$ before the $1/\eta$ correction becomes relevant, i.e. $m^*\approx \log_3\eta$. Hence, for $\eta\approx 10^6$ we expect that the exponential scaling holds up to $m^*\approx 10$.

\section{Conclusions}\label{s:conc}
In this article we have reported a quantum metrological scheme that yields a quantum Fisher information that scales exponentially with the protocol duration time $T$. Such scheme is rooted in the non-adiabaticity of a cycle in the control parameter reaching a quantum critical point in fully-connected models. In one cycle, the state acquires squeezing, which can be amplified under a suitable choice of parameters by subsequent cycles. As we show, after $m$ cycles of a duration $2\tau$ each, the quantum Fisher information scales as $I_\omega\propto 3^{T/\tau}$. This scaling is well captured by the recent bound put forward in Ref.~\cite{Garbe21}, which in turn allows us to find approximated expressions for the quantum Fisher information. We discuss the potential deviations to this ideal scenario, such as finite-size effects or the effect of decoherence mechanisms. Supported by numerical simulations, we find that the exponential precision is robust against decoherence effects. In addition, as we argue, finite-size effects pose a limit to the maximum number of cycles that can be performed before the exponential scaling breaks down.

Our results, together with those reported recently by Gietka {\em et al.}~\cite{Gietka21}, highlight that systems featuring a quantum phase transition are a valuable resource in quantum metrology as they can yield an exponential advantage for parameter estimation.


\begin{thebibliography}{77}%
\makeatletter
\providecommand \@ifxundefined [1]{%
 \@ifx{#1\undefined}
}%
\providecommand \@ifnum [1]{%
 \ifnum #1\expandafter \@firstoftwo
 \else \expandafter \@secondoftwo
 \fi
}%
\providecommand \@ifx [1]{%
 \ifx #1\expandafter \@firstoftwo
 \else \expandafter \@secondoftwo
 \fi
}%
\providecommand \natexlab [1]{#1}%
\providecommand \enquote  [1]{``#1''}%
\providecommand \bibnamefont  [1]{#1}%
\providecommand \bibfnamefont [1]{#1}%
\providecommand \citenamefont [1]{#1}%
\providecommand \href@noop [0]{\@secondoftwo}%
\providecommand \href [0]{\begingroup \@sanitize@url \@href}%
\providecommand \@href[1]{\@@startlink{#1}\@@href}%
\providecommand \@@href[1]{\endgroup#1\@@endlink}%
\providecommand \@sanitize@url [0]{\catcode `\\12\catcode `\$12\catcode
  `\&12\catcode `\#12\catcode `\^12\catcode `\_12\catcode `\%12\relax}%
\providecommand \@@startlink[1]{}%
\providecommand \@@endlink[0]{}%
\providecommand \url  [0]{\begingroup\@sanitize@url \@url }%
\providecommand \@url [1]{\endgroup\@href {#1}{\urlprefix }}%
\providecommand \urlprefix  [0]{URL }%
\providecommand \Eprint [0]{\href }%
\providecommand \doibase [0]{http://dx.doi.org/}%
\providecommand \selectlanguage [0]{\@gobble}%
\providecommand \bibinfo  [0]{\@secondoftwo}%
\providecommand \bibfield  [0]{\@secondoftwo}%
\providecommand \translation [1]{[#1]}%
\providecommand \BibitemOpen [0]{}%
\providecommand \bibitemStop [0]{}%
\providecommand \bibitemNoStop [0]{.\EOS\space}%
\providecommand \EOS [0]{\spacefactor3000\relax}%
\providecommand \BibitemShut  [1]{\csname bibitem#1\endcsname}%
\let\auto@bib@innerbib\@empty
\bibitem [{\citenamefont {Garbe}\ \emph {et~al.}(2021)\citenamefont {Garbe},
  \citenamefont {Abah}, \citenamefont {Felicetti},\ and\ \citenamefont
  {Puebla}}]{Garbe21}%
  \BibitemOpen
  \bibfield  {author} {\bibinfo {author} {\bibfnamefont {L.}~\bibnamefont
  {Garbe}}, \bibinfo {author} {\bibfnamefont {O.}~\bibnamefont {Abah}},
  \bibinfo {author} {\bibfnamefont {S.}~\bibnamefont {Felicetti}}, \ and\
  \bibinfo {author} {\bibfnamefont {R.}~\bibnamefont {Puebla}},\ }\href
  {https://arxiv.org/abs/2110.04144} {\bibfield  {journal} {\bibinfo  {journal}
  {arXiv:2110.04144}\ } (\bibinfo {year} {2021})}\BibitemShut {NoStop}%
\bibitem [{\citenamefont {Giovannetti}\ \emph {et~al.}(2004)\citenamefont
  {Giovannetti}, \citenamefont {Lloyd},\ and\ \citenamefont
  {Maccone}}]{Giovannetti:04}%
  \BibitemOpen
  \bibfield  {author} {\bibinfo {author} {\bibfnamefont {V.}~\bibnamefont
  {Giovannetti}}, \bibinfo {author} {\bibfnamefont {S.}~\bibnamefont {Lloyd}},
  \ and\ \bibinfo {author} {\bibfnamefont {L.}~\bibnamefont {Maccone}},\ }\href
  {\doibase 10.1126/science.1104149} {\bibfield  {journal} {\bibinfo  {journal}
  {Science}\ }\textbf {\bibinfo {volume} {306}},\ \bibinfo {pages} {1330}
  (\bibinfo {year} {2004})}\BibitemShut {NoStop}%
\bibitem [{\citenamefont {Giovannetti}\ \emph {et~al.}(2006)\citenamefont
  {Giovannetti}, \citenamefont {Lloyd},\ and\ \citenamefont
  {Maccone}}]{Giovannetti:06}%
  \BibitemOpen
  \bibfield  {author} {\bibinfo {author} {\bibfnamefont {V.}~\bibnamefont
  {Giovannetti}}, \bibinfo {author} {\bibfnamefont {S.}~\bibnamefont {Lloyd}},
  \ and\ \bibinfo {author} {\bibfnamefont {L.}~\bibnamefont {Maccone}},\ }\href
  {\doibase 10.1103/PhysRevLett.96.010401} {\bibfield  {journal} {\bibinfo
  {journal} {Phys. Rev. Lett.}\ }\textbf {\bibinfo {volume} {96}},\ \bibinfo
  {pages} {010401} (\bibinfo {year} {2006})}\BibitemShut {NoStop}%
\bibitem [{\citenamefont {Paris}(2009)}]{Paris:11}%
  \BibitemOpen
  \bibfield  {author} {\bibinfo {author} {\bibfnamefont {M.~G.~A.}\
  \bibnamefont {Paris}},\ }\href {\doibase 10.1142/S0219749909004839}
  {\bibfield  {journal} {\bibinfo  {journal} {Int. J. Quantum Inf.}\ }\textbf
  {\bibinfo {volume} {7}},\ \bibinfo {pages} {125} (\bibinfo {year}
  {2009})}\BibitemShut {NoStop}%
\bibitem [{\citenamefont {Pezz\`e}\ \emph {et~al.}(2018)\citenamefont
  {Pezz\`e}, \citenamefont {Smerzi}, \citenamefont {Oberthaler}, \citenamefont
  {Schmied},\ and\ \citenamefont {Treutlein}}]{Pezze:18}%
  \BibitemOpen
  \bibfield  {author} {\bibinfo {author} {\bibfnamefont {L.}~\bibnamefont
  {Pezz\`e}}, \bibinfo {author} {\bibfnamefont {A.}~\bibnamefont {Smerzi}},
  \bibinfo {author} {\bibfnamefont {M.~K.}\ \bibnamefont {Oberthaler}},
  \bibinfo {author} {\bibfnamefont {R.}~\bibnamefont {Schmied}}, \ and\
  \bibinfo {author} {\bibfnamefont {P.}~\bibnamefont {Treutlein}},\ }\href
  {\doibase 10.1103/RevModPhys.90.035005} {\bibfield  {journal} {\bibinfo
  {journal} {Rev. Mod. Phys.}\ }\textbf {\bibinfo {volume} {90}},\ \bibinfo
  {pages} {035005} (\bibinfo {year} {2018})}\BibitemShut {NoStop}%
\bibitem [{\citenamefont {Dowling}\ and\ \citenamefont
  {Milburn}(2003)}]{Dowling:03}%
  \BibitemOpen
  \bibfield  {author} {\bibinfo {author} {\bibfnamefont {J.~P.}\ \bibnamefont
  {Dowling}}\ and\ \bibinfo {author} {\bibfnamefont {G.~J.}\ \bibnamefont
  {Milburn}},\ }\href {\doibase 10.1098/rsta.2003.1227} {\bibfield  {journal}
  {\bibinfo  {journal} {Phil. Trans. R. Soc. A}\ }\textbf {\bibinfo {volume}
  {361}},\ \bibinfo {pages} {1655} (\bibinfo {year} {2003})}\BibitemShut
  {NoStop}%
\bibitem [{\citenamefont {Braunstein}\ and\ \citenamefont
  {Caves}(1994)}]{Braunstein:94}%
  \BibitemOpen
  \bibfield  {author} {\bibinfo {author} {\bibfnamefont {S.~L.}\ \bibnamefont
  {Braunstein}}\ and\ \bibinfo {author} {\bibfnamefont {C.~M.}\ \bibnamefont
  {Caves}},\ }\href {\doibase 10.1103/PhysRevLett.72.3439} {\bibfield
  {journal} {\bibinfo  {journal} {Phys. Rev. Lett.}\ }\textbf {\bibinfo
  {volume} {72}},\ \bibinfo {pages} {3439} (\bibinfo {year}
  {1994})}\BibitemShut {NoStop}%
\bibitem [{\citenamefont {Boixo}\ \emph {et~al.}(2007)\citenamefont {Boixo},
  \citenamefont {Flammia}, \citenamefont {Caves},\ and\ \citenamefont
  {Geremia}}]{Boixo:07}%
  \BibitemOpen
  \bibfield  {author} {\bibinfo {author} {\bibfnamefont {S.}~\bibnamefont
  {Boixo}}, \bibinfo {author} {\bibfnamefont {S.~T.}\ \bibnamefont {Flammia}},
  \bibinfo {author} {\bibfnamefont {C.~M.}\ \bibnamefont {Caves}}, \ and\
  \bibinfo {author} {\bibfnamefont {J.}~\bibnamefont {Geremia}},\ }\href
  {\doibase 10.1103/PhysRevLett.98.090401} {\bibfield  {journal} {\bibinfo
  {journal} {Phys. Rev. Lett.}\ }\textbf {\bibinfo {volume} {98}},\ \bibinfo
  {pages} {090401} (\bibinfo {year} {2007})}\BibitemShut {NoStop}%
\bibitem [{\citenamefont {Roy}\ and\ \citenamefont
  {Braunstein}(2008)}]{Roy:08}%
  \BibitemOpen
  \bibfield  {author} {\bibinfo {author} {\bibfnamefont {S.~M.}\ \bibnamefont
  {Roy}}\ and\ \bibinfo {author} {\bibfnamefont {S.~L.}\ \bibnamefont
  {Braunstein}},\ }\href {\doibase 10.1103/PhysRevLett.100.220501} {\bibfield
  {journal} {\bibinfo  {journal} {Phys. Rev. Lett.}\ }\textbf {\bibinfo
  {volume} {100}},\ \bibinfo {pages} {220501} (\bibinfo {year}
  {2008})}\BibitemShut {NoStop}%
\bibitem [{\citenamefont {Pang}\ and\ \citenamefont {Jordan}(2017)}]{Pang:17}%
  \BibitemOpen
  \bibfield  {author} {\bibinfo {author} {\bibfnamefont {S.}~\bibnamefont
  {Pang}}\ and\ \bibinfo {author} {\bibfnamefont {A.~N.}\ \bibnamefont
  {Jordan}},\ }\href {\doibase 10.1038/ncomms14695} {\bibfield  {journal}
  {\bibinfo  {journal} {Nat. Commun.}\ }\textbf {\bibinfo {volume} {8}},\
  \bibinfo {pages} {14695} (\bibinfo {year} {2017})}\BibitemShut {NoStop}%
\bibitem [{\citenamefont {Sachdev}(2011)}]{Sachdev}%
  \BibitemOpen
  \bibfield  {author} {\bibinfo {author} {\bibfnamefont {S.}~\bibnamefont
  {Sachdev}},\ }\href@noop {} {\emph {\bibinfo {title} {{Quantum phase
  transitions}}}},\ \bibinfo {edition} {2nd}\ ed.\ (\bibinfo  {publisher}
  {Cambridge University Press},\ \bibinfo {address} {Cambridge, UK},\ \bibinfo
  {year} {2011})\BibitemShut {NoStop}%
\bibitem [{\citenamefont {Zanardi}\ \emph {et~al.}(2008)\citenamefont
  {Zanardi}, \citenamefont {Paris},\ and\ \citenamefont
  {Campos~Venuti}}]{Zanardi:08}%
  \BibitemOpen
  \bibfield  {author} {\bibinfo {author} {\bibfnamefont {P.}~\bibnamefont
  {Zanardi}}, \bibinfo {author} {\bibfnamefont {M.~G.~A.}\ \bibnamefont
  {Paris}}, \ and\ \bibinfo {author} {\bibfnamefont {L.}~\bibnamefont
  {Campos~Venuti}},\ }\href {\doibase 10.1103/PhysRevA.78.042105} {\bibfield
  {journal} {\bibinfo  {journal} {Phys. Rev. A}\ }\textbf {\bibinfo {volume}
  {78}},\ \bibinfo {pages} {042105} (\bibinfo {year} {2008})}\BibitemShut
  {NoStop}%
\bibitem [{\citenamefont {Bina}\ \emph {et~al.}(2016)\citenamefont {Bina},
  \citenamefont {Amelio},\ and\ \citenamefont {Paris}}]{bina_dicke_2016}%
  \BibitemOpen
  \bibfield  {author} {\bibinfo {author} {\bibfnamefont {M.}~\bibnamefont
  {Bina}}, \bibinfo {author} {\bibfnamefont {I.}~\bibnamefont {Amelio}}, \ and\
  \bibinfo {author} {\bibfnamefont {M.~G.~A.}\ \bibnamefont {Paris}},\ }\href
  {\doibase 10.1103/PhysRevE.93.052118} {\bibfield  {journal} {\bibinfo
  {journal} {Phys. Rev. E}\ }\textbf {\bibinfo {volume} {93}},\ \bibinfo
  {pages} {052118} (\bibinfo {year} {2016})}\BibitemShut {NoStop}%
\bibitem [{\citenamefont {Fern\'andez-Lorenzo}\ and\ \citenamefont
  {Porras}(2017)}]{FernandezLorenzo:17}%
  \BibitemOpen
  \bibfield  {author} {\bibinfo {author} {\bibfnamefont {S.}~\bibnamefont
  {Fern\'andez-Lorenzo}}\ and\ \bibinfo {author} {\bibfnamefont
  {D.}~\bibnamefont {Porras}},\ }\href {\doibase 10.1103/PhysRevA.96.013817}
  {\bibfield  {journal} {\bibinfo  {journal} {Phys. Rev. A}\ }\textbf {\bibinfo
  {volume} {96}},\ \bibinfo {pages} {013817} (\bibinfo {year}
  {2017})}\BibitemShut {NoStop}%
\bibitem [{\citenamefont {Rams}\ \emph {et~al.}(2018)\citenamefont {Rams},
  \citenamefont {Sierant}, \citenamefont {Dutta}, \citenamefont {Horodecki},\
  and\ \citenamefont {Zakrzewski}}]{Rams:18}%
  \BibitemOpen
  \bibfield  {author} {\bibinfo {author} {\bibfnamefont {M.~M.}\ \bibnamefont
  {Rams}}, \bibinfo {author} {\bibfnamefont {P.}~\bibnamefont {Sierant}},
  \bibinfo {author} {\bibfnamefont {O.}~\bibnamefont {Dutta}}, \bibinfo
  {author} {\bibfnamefont {P.}~\bibnamefont {Horodecki}}, \ and\ \bibinfo
  {author} {\bibfnamefont {J.}~\bibnamefont {Zakrzewski}},\ }\href {\doibase
  10.1103/PhysRevX.8.021022} {\bibfield  {journal} {\bibinfo  {journal} {Phys.
  Rev. X}\ }\textbf {\bibinfo {volume} {8}},\ \bibinfo {pages} {021022}
  (\bibinfo {year} {2018})}\BibitemShut {NoStop}%
\bibitem [{\citenamefont {Garbe}\ \emph {et~al.}(2020)\citenamefont {Garbe},
  \citenamefont {Bina}, \citenamefont {Keller}, \citenamefont {Paris},\ and\
  \citenamefont {Felicetti}}]{Garbe:20}%
  \BibitemOpen
  \bibfield  {author} {\bibinfo {author} {\bibfnamefont {L.}~\bibnamefont
  {Garbe}}, \bibinfo {author} {\bibfnamefont {M.}~\bibnamefont {Bina}},
  \bibinfo {author} {\bibfnamefont {A.}~\bibnamefont {Keller}}, \bibinfo
  {author} {\bibfnamefont {M.~G.~A.}\ \bibnamefont {Paris}}, \ and\ \bibinfo
  {author} {\bibfnamefont {S.}~\bibnamefont {Felicetti}},\ }\href {\doibase
  10.1103/PhysRevLett.124.120504} {\bibfield  {journal} {\bibinfo  {journal}
  {Phys. Rev. Lett.}\ }\textbf {\bibinfo {volume} {124}},\ \bibinfo {pages}
  {120504} (\bibinfo {year} {2020})}\BibitemShut {NoStop}%
\bibitem [{\citenamefont {Ivanov}(2020)}]{Ivanov:20}%
  \BibitemOpen
  \bibfield  {author} {\bibinfo {author} {\bibfnamefont {P.~A.}\ \bibnamefont
  {Ivanov}},\ }\href {\doibase 10.1103/PhysRevA.102.052611} {\bibfield
  {journal} {\bibinfo  {journal} {Phys. Rev. A}\ }\textbf {\bibinfo {volume}
  {102}},\ \bibinfo {pages} {052611} (\bibinfo {year} {2020})}\BibitemShut
  {NoStop}%
\bibitem [{\citenamefont {Chu}\ \emph {et~al.}(2021)\citenamefont {Chu},
  \citenamefont {Zhang}, \citenamefont {Yu},\ and\ \citenamefont
  {Cai}}]{Chu:21}%
  \BibitemOpen
  \bibfield  {author} {\bibinfo {author} {\bibfnamefont {Y.}~\bibnamefont
  {Chu}}, \bibinfo {author} {\bibfnamefont {S.}~\bibnamefont {Zhang}}, \bibinfo
  {author} {\bibfnamefont {B.}~\bibnamefont {Yu}}, \ and\ \bibinfo {author}
  {\bibfnamefont {J.}~\bibnamefont {Cai}},\ }\href {\doibase
  10.1103/PhysRevLett.126.010502} {\bibfield  {journal} {\bibinfo  {journal}
  {Phys. Rev. Lett.}\ }\textbf {\bibinfo {volume} {126}},\ \bibinfo {pages}
  {010502} (\bibinfo {year} {2021})}\BibitemShut {NoStop}%
\bibitem [{\citenamefont {Candia}\ \emph {et~al.}(2021)\citenamefont {Candia},
  \citenamefont {Minganti}, \citenamefont {Petrovnin}, \citenamefont
  {Paraoanu},\ and\ \citenamefont {Felicetti}}]{DiCandia:21}%
  \BibitemOpen
  \bibfield  {author} {\bibinfo {author} {\bibfnamefont {R.~D.}\ \bibnamefont
  {Candia}}, \bibinfo {author} {\bibfnamefont {F.}~\bibnamefont {Minganti}},
  \bibinfo {author} {\bibfnamefont {K.~V.}\ \bibnamefont {Petrovnin}}, \bibinfo
  {author} {\bibfnamefont {G.~S.}\ \bibnamefont {Paraoanu}}, \ and\ \bibinfo
  {author} {\bibfnamefont {S.}~\bibnamefont {Felicetti}},\ }\href
  {https://arxiv.org/abs/2107.04503} {\bibfield  {journal} {\bibinfo  {journal}
  {arXiv:2107.04503}\ } (\bibinfo {year} {2021})}\BibitemShut {NoStop}%
\bibitem [{\citenamefont {Ilias}\ \emph {et~al.}(2021)\citenamefont {Ilias},
  \citenamefont {Yang}, \citenamefont {Huelga},\ and\ \citenamefont
  {Plenio}}]{Ilias:21}%
  \BibitemOpen
  \bibfield  {author} {\bibinfo {author} {\bibfnamefont {T.}~\bibnamefont
  {Ilias}}, \bibinfo {author} {\bibfnamefont {D.}~\bibnamefont {Yang}},
  \bibinfo {author} {\bibfnamefont {S.~F.}\ \bibnamefont {Huelga}}, \ and\
  \bibinfo {author} {\bibfnamefont {M.~B.}\ \bibnamefont {Plenio}},\ }\href
  {https://arxiv.org/abs/2108.06349} {\bibfield  {journal} {\bibinfo  {journal}
  {arXiv:2108.06349}\ } (\bibinfo {year} {2021})}\BibitemShut {NoStop}%
\bibitem [{\citenamefont {Hu}\ \emph {et~al.}(2021)\citenamefont {Hu},
  \citenamefont {Huang}, \citenamefont {Huang}, \citenamefont {Xie},\ and\
  \citenamefont {Liao}}]{hu2021}%
  \BibitemOpen
  \bibfield  {author} {\bibinfo {author} {\bibfnamefont {Y.}~\bibnamefont
  {Hu}}, \bibinfo {author} {\bibfnamefont {J.}~\bibnamefont {Huang}}, \bibinfo
  {author} {\bibfnamefont {J.-F.}\ \bibnamefont {Huang}}, \bibinfo {author}
  {\bibfnamefont {Q.-T.}\ \bibnamefont {Xie}}, \ and\ \bibinfo {author}
  {\bibfnamefont {J.-Q.}\ \bibnamefont {Liao}},\ }\href
  {https://arxiv.org/abs/2101.01504} {\bibfield  {journal} {\bibinfo  {journal}
  {arXiv:2101.01504}\ } (\bibinfo {year} {2021})}\BibitemShut {NoStop}%
\bibitem [{\citenamefont {Fallani}\ \emph {et~al.}(2021)\citenamefont
  {Fallani}, \citenamefont {Rossi}, \citenamefont {Tamascelli},\ and\
  \citenamefont {Genoni}}]{fallani2021learning}%
  \BibitemOpen
  \bibfield  {author} {\bibinfo {author} {\bibfnamefont {A.}~\bibnamefont
  {Fallani}}, \bibinfo {author} {\bibfnamefont {M.~A.~C.}\ \bibnamefont
  {Rossi}}, \bibinfo {author} {\bibfnamefont {D.}~\bibnamefont {Tamascelli}}, \
  and\ \bibinfo {author} {\bibfnamefont {M.~G.}\ \bibnamefont {Genoni}},\
  }\href {https://arxiv.org/abs/2110.15080} {\bibfield  {journal} {\bibinfo
  {journal} {arXiv:2110.15080}\ } (\bibinfo {year} {2021})}\BibitemShut
  {NoStop}%
\bibitem [{\citenamefont {Xie}\ \emph {et~al.}(2021)\citenamefont {Xie},
  \citenamefont {Xu},\ and\ \citenamefont {Wang}}]{xie2021}%
  \BibitemOpen
  \bibfield  {author} {\bibinfo {author} {\bibfnamefont {D.}~\bibnamefont
  {Xie}}, \bibinfo {author} {\bibfnamefont {C.}~\bibnamefont {Xu}}, \ and\
  \bibinfo {author} {\bibfnamefont {A.~M.}\ \bibnamefont {Wang}},\ }\href
  {https://arxiv.org/abs/2101.01504} {\bibfield  {journal} {\bibinfo  {journal}
  {arXiv:2105.12906}\ } (\bibinfo {year} {2021})}\BibitemShut {NoStop}%
\bibitem [{\citenamefont {Niezgoda}\ and\ \citenamefont
  {Chwede\ifmmode~\acute{n}\else \'{n}\fi{}czuk}(2021)}]{Niezgoda2021}%
  \BibitemOpen
  \bibfield  {author} {\bibinfo {author} {\bibfnamefont {A.}~\bibnamefont
  {Niezgoda}}\ and\ \bibinfo {author} {\bibfnamefont {J.}~\bibnamefont
  {Chwede\ifmmode~\acute{n}\else \'{n}\fi{}czuk}},\ }\href {\doibase
  10.1103/PhysRevLett.126.210506} {\bibfield  {journal} {\bibinfo  {journal}
  {Phys. Rev. Lett.}\ }\textbf {\bibinfo {volume} {126}},\ \bibinfo {pages}
  {210506} (\bibinfo {year} {2021})}\BibitemShut {NoStop}%
\bibitem [{\citenamefont {Gietka}\ \emph {et~al.}(2021)\citenamefont {Gietka},
  \citenamefont {Ruks},\ and\ \citenamefont {Busch}}]{Gietka21}%
  \BibitemOpen
  \bibfield  {author} {\bibinfo {author} {\bibfnamefont {K.}~\bibnamefont
  {Gietka}}, \bibinfo {author} {\bibfnamefont {L.}~\bibnamefont {Ruks}}, \ and\
  \bibinfo {author} {\bibfnamefont {T.}~\bibnamefont {Busch}},\ }\href
  {https://arxiv.org/abs/2110.04048} {\bibfield  {journal} {\bibinfo  {journal}
  {arXiv:2110.04048}\ } (\bibinfo {year} {2021})}\BibitemShut {NoStop}%
\bibitem [{\citenamefont {Montenegro}\ \emph {et~al.}(2021)\citenamefont
  {Montenegro}, \citenamefont {Mishra},\ and\ \citenamefont
  {Bayat}}]{Montenegro2021}%
  \BibitemOpen
  \bibfield  {author} {\bibinfo {author} {\bibfnamefont {V.}~\bibnamefont
  {Montenegro}}, \bibinfo {author} {\bibfnamefont {U.}~\bibnamefont {Mishra}},
  \ and\ \bibinfo {author} {\bibfnamefont {A.}~\bibnamefont {Bayat}},\ }\href
  {\doibase 10.1103/PhysRevLett.126.200501} {\bibfield  {journal} {\bibinfo
  {journal} {Phys. Rev. Lett.}\ }\textbf {\bibinfo {volume} {126}},\ \bibinfo
  {pages} {200501} (\bibinfo {year} {2021})}\BibitemShut {NoStop}%
\bibitem [{\citenamefont {Ivanov}(2021)}]{ivanov2021enhanced}%
  \BibitemOpen
  \bibfield  {author} {\bibinfo {author} {\bibfnamefont {P.~A.}\ \bibnamefont
  {Ivanov}},\ }\href@noop {} {\bibfield  {journal} {\bibinfo  {journal}
  {Entropy}\ }\textbf {\bibinfo {volume} {23}},\ \bibinfo {pages} {1333}
  (\bibinfo {year} {2021})}\BibitemShut {NoStop}%
\bibitem [{\citenamefont {Gietka}(2021)}]{gietka2021squeezing}%
  \BibitemOpen
  \bibfield  {author} {\bibinfo {author} {\bibfnamefont {K.}~\bibnamefont
  {Gietka}},\ }\href {https://arxiv.org/abs/2111.12206} {\bibfield  {journal}
  {\bibinfo  {journal} {arXiv:2111.12206}\ } (\bibinfo {year}
  {2021})}\BibitemShut {NoStop}%
\bibitem [{\citenamefont {Gietka}\ and\ \citenamefont
  {Busch}(2021)}]{Gietka_inverted}%
  \BibitemOpen
  \bibfield  {author} {\bibinfo {author} {\bibfnamefont {K.}~\bibnamefont
  {Gietka}}\ and\ \bibinfo {author} {\bibfnamefont {T.}~\bibnamefont {Busch}},\
  }\href {\doibase 10.1103/PhysRevE.104.034132} {\bibfield  {journal} {\bibinfo
   {journal} {Phys. Rev. E}\ }\textbf {\bibinfo {volume} {104}},\ \bibinfo
  {pages} {034132} (\bibinfo {year} {2021})}\BibitemShut {NoStop}%
\bibitem [{\citenamefont {Weiss}\ \emph {et~al.}(2021)\citenamefont {Weiss},
  \citenamefont {Roda-Llordes}, \citenamefont {Torrontegui}, \citenamefont
  {Aspelmeyer},\ and\ \citenamefont {Romero-Isart}}]{Weiss21}%
  \BibitemOpen
  \bibfield  {author} {\bibinfo {author} {\bibfnamefont {T.}~\bibnamefont
  {Weiss}}, \bibinfo {author} {\bibfnamefont {M.}~\bibnamefont {Roda-Llordes}},
  \bibinfo {author} {\bibfnamefont {E.}~\bibnamefont {Torrontegui}}, \bibinfo
  {author} {\bibfnamefont {M.}~\bibnamefont {Aspelmeyer}}, \ and\ \bibinfo
  {author} {\bibfnamefont {O.}~\bibnamefont {Romero-Isart}},\ }\href {\doibase
  10.1103/PhysRevLett.127.023601} {\bibfield  {journal} {\bibinfo  {journal}
  {Phys. Rev. Lett.}\ }\textbf {\bibinfo {volume} {127}},\ \bibinfo {pages}
  {023601} (\bibinfo {year} {2021})}\BibitemShut {NoStop}%
\bibitem [{\citenamefont {Yamamoto}\ \emph {et~al.}(2021)\citenamefont
  {Yamamoto}, \citenamefont {Endo}, \citenamefont {Hakoshima}, \citenamefont
  {Matsuzaki},\ and\ \citenamefont {Tokunaga}}]{Yamamoto:21}%
  \BibitemOpen
  \bibfield  {author} {\bibinfo {author} {\bibfnamefont {K.}~\bibnamefont
  {Yamamoto}}, \bibinfo {author} {\bibfnamefont {S.}~\bibnamefont {Endo}},
  \bibinfo {author} {\bibfnamefont {H.}~\bibnamefont {Hakoshima}}, \bibinfo
  {author} {\bibfnamefont {Y.}~\bibnamefont {Matsuzaki}}, \ and\ \bibinfo
  {author} {\bibfnamefont {Y.}~\bibnamefont {Tokunaga}},\ }\href
  {https://arxiv.org/abs/2112.01850} {\bibfield  {journal} {\bibinfo  {journal}
  {arXiv:2112.01850}\ } (\bibinfo {year} {2021})}\BibitemShut {NoStop}%
\bibitem [{\citenamefont {Cosco}\ \emph {et~al.}(2021)\citenamefont {Cosco},
  \citenamefont {Pedernales},\ and\ \citenamefont {Plenio}}]{Cosco:21}%
  \BibitemOpen
  \bibfield  {author} {\bibinfo {author} {\bibfnamefont {F.}~\bibnamefont
  {Cosco}}, \bibinfo {author} {\bibfnamefont {J.~S.}\ \bibnamefont
  {Pedernales}}, \ and\ \bibinfo {author} {\bibfnamefont {M.~B.}\ \bibnamefont
  {Plenio}},\ }\href {\doibase 10.1103/PhysRevA.103.L061501} {\bibfield
  {journal} {\bibinfo  {journal} {Phys. Rev. A}\ }\textbf {\bibinfo {volume}
  {103}},\ \bibinfo {pages} {L061501} (\bibinfo {year} {2021})}\BibitemShut
  {NoStop}%
\bibitem [{\citenamefont {Lipkin}\ \emph {et~al.}(1965)\citenamefont {Lipkin},
  \citenamefont {Meshkov},\ and\ \citenamefont {Glick}}]{Lipkin:65}%
  \BibitemOpen
  \bibfield  {author} {\bibinfo {author} {\bibfnamefont {H.~J.}\ \bibnamefont
  {Lipkin}}, \bibinfo {author} {\bibfnamefont {N.}~\bibnamefont {Meshkov}}, \
  and\ \bibinfo {author} {\bibfnamefont {A.}~\bibnamefont {Glick}},\ }\href
  {\doibase http://dx.doi.org/10.1016/0029-5582(65)90862-X} {\bibfield
  {journal} {\bibinfo  {journal} {Nucl. Phys.}\ }\textbf {\bibinfo {volume}
  {62}},\ \bibinfo {pages} {188} (\bibinfo {year} {1965})}\BibitemShut
  {NoStop}%
\bibitem [{\citenamefont {Dicke}(1954)}]{Dicke:54}%
  \BibitemOpen
  \bibfield  {author} {\bibinfo {author} {\bibfnamefont {R.~H.}\ \bibnamefont
  {Dicke}},\ }\href {\doibase 10.1103/PhysRev.93.99} {\bibfield  {journal}
  {\bibinfo  {journal} {Phys. Rev.}\ }\textbf {\bibinfo {volume} {93}},\
  \bibinfo {pages} {99} (\bibinfo {year} {1954})}\BibitemShut {NoStop}%
\bibitem [{\citenamefont {Ashhab}(2013)}]{Ashhab:13}%
  \BibitemOpen
  \bibfield  {author} {\bibinfo {author} {\bibfnamefont {S.}~\bibnamefont
  {Ashhab}},\ }\href {\doibase 10.1103/PhysRevA.87.013826} {\bibfield
  {journal} {\bibinfo  {journal} {Phys. Rev. A}\ }\textbf {\bibinfo {volume}
  {87}},\ \bibinfo {pages} {013826} (\bibinfo {year} {2013})}\BibitemShut
  {NoStop}%
\bibitem [{\citenamefont {Bakemeier}\ \emph {et~al.}(2012)\citenamefont
  {Bakemeier}, \citenamefont {Alvermann},\ and\ \citenamefont
  {Fehske}}]{Bakemeier:12}%
  \BibitemOpen
  \bibfield  {author} {\bibinfo {author} {\bibfnamefont {L.}~\bibnamefont
  {Bakemeier}}, \bibinfo {author} {\bibfnamefont {A.}~\bibnamefont
  {Alvermann}}, \ and\ \bibinfo {author} {\bibfnamefont {H.}~\bibnamefont
  {Fehske}},\ }\href {\doibase 10.1103/PhysRevA.85.043821} {\bibfield
  {journal} {\bibinfo  {journal} {Phys. Rev. A}\ }\textbf {\bibinfo {volume}
  {85}},\ \bibinfo {pages} {043821} (\bibinfo {year} {2012})}\BibitemShut
  {NoStop}%
\bibitem [{\citenamefont {Hwang}\ \emph {et~al.}(2015)\citenamefont {Hwang},
  \citenamefont {Puebla},\ and\ \citenamefont {Plenio}}]{Hwang:15}%
  \BibitemOpen
  \bibfield  {author} {\bibinfo {author} {\bibfnamefont {M.-J.}\ \bibnamefont
  {Hwang}}, \bibinfo {author} {\bibfnamefont {R.}~\bibnamefont {Puebla}}, \
  and\ \bibinfo {author} {\bibfnamefont {M.~B.}\ \bibnamefont {Plenio}},\
  }\href {\doibase 10.1103/PhysRevLett.115.180404} {\bibfield  {journal}
  {\bibinfo  {journal} {Phys. Rev. Lett.}\ }\textbf {\bibinfo {volume} {115}},\
  \bibinfo {pages} {180404} (\bibinfo {year} {2015})}\BibitemShut {NoStop}%
\bibitem [{\citenamefont {Puebla}\ \emph {et~al.}(2016)\citenamefont {Puebla},
  \citenamefont {Hwang},\ and\ \citenamefont {Plenio}}]{Puebla:16}%
  \BibitemOpen
  \bibfield  {author} {\bibinfo {author} {\bibfnamefont {R.}~\bibnamefont
  {Puebla}}, \bibinfo {author} {\bibfnamefont {M.-J.}\ \bibnamefont {Hwang}}, \
  and\ \bibinfo {author} {\bibfnamefont {M.~B.}\ \bibnamefont {Plenio}},\
  }\href {\doibase 10.1103/PhysRevA.94.023835} {\bibfield  {journal} {\bibinfo
  {journal} {Phys. Rev. A}\ }\textbf {\bibinfo {volume} {94}},\ \bibinfo
  {pages} {023835} (\bibinfo {year} {2016})}\BibitemShut {NoStop}%
\bibitem [{\citenamefont {Ribeiro}\ \emph {et~al.}(2007)\citenamefont
  {Ribeiro}, \citenamefont {Vidal},\ and\ \citenamefont
  {Mosseri}}]{Ribeiro:07}%
  \BibitemOpen
  \bibfield  {author} {\bibinfo {author} {\bibfnamefont {P.}~\bibnamefont
  {Ribeiro}}, \bibinfo {author} {\bibfnamefont {J.}~\bibnamefont {Vidal}}, \
  and\ \bibinfo {author} {\bibfnamefont {R.}~\bibnamefont {Mosseri}},\ }\href
  {\doibase 10.1103/PhysRevLett.99.050402} {\bibfield  {journal} {\bibinfo
  {journal} {Phys. Rev. Lett.}\ }\textbf {\bibinfo {volume} {99}},\ \bibinfo
  {pages} {050402} (\bibinfo {year} {2007})}\BibitemShut {NoStop}%
\bibitem [{\citenamefont {Ribeiro}\ \emph {et~al.}(2008)\citenamefont
  {Ribeiro}, \citenamefont {Vidal},\ and\ \citenamefont
  {Mosseri}}]{Ribeiro:08}%
  \BibitemOpen
  \bibfield  {author} {\bibinfo {author} {\bibfnamefont {P.}~\bibnamefont
  {Ribeiro}}, \bibinfo {author} {\bibfnamefont {J.}~\bibnamefont {Vidal}}, \
  and\ \bibinfo {author} {\bibfnamefont {R.}~\bibnamefont {Mosseri}},\ }\href
  {\doibase 10.1103/PhysRevE.78.021106} {\bibfield  {journal} {\bibinfo
  {journal} {Phys. Rev. E}\ }\textbf {\bibinfo {volume} {78}},\ \bibinfo
  {pages} {021106} (\bibinfo {year} {2008})}\BibitemShut {NoStop}%
\bibitem [{\citenamefont {Emary}\ and\ \citenamefont
  {Brandes}(2003{\natexlab{a}})}]{Emary:03prl}%
  \BibitemOpen
  \bibfield  {author} {\bibinfo {author} {\bibfnamefont {C.}~\bibnamefont
  {Emary}}\ and\ \bibinfo {author} {\bibfnamefont {T.}~\bibnamefont
  {Brandes}},\ }\href {\doibase 10.1103/PhysRevLett.90.044101} {\bibfield
  {journal} {\bibinfo  {journal} {Phys. Rev. Lett.}\ }\textbf {\bibinfo
  {volume} {90}},\ \bibinfo {pages} {044101} (\bibinfo {year}
  {2003}{\natexlab{a}})}\BibitemShut {NoStop}%
\bibitem [{\citenamefont {Emary}\ and\ \citenamefont
  {Brandes}(2003{\natexlab{b}})}]{Emary:03}%
  \BibitemOpen
  \bibfield  {author} {\bibinfo {author} {\bibfnamefont {C.}~\bibnamefont
  {Emary}}\ and\ \bibinfo {author} {\bibfnamefont {T.}~\bibnamefont
  {Brandes}},\ }\href {\doibase 10.1103/PhysRevE.67.066203} {\bibfield
  {journal} {\bibinfo  {journal} {Phys. Rev. E}\ }\textbf {\bibinfo {volume}
  {67}},\ \bibinfo {pages} {066203} (\bibinfo {year}
  {2003}{\natexlab{b}})}\BibitemShut {NoStop}%
\bibitem [{\citenamefont {Lambert}\ \emph {et~al.}(2004)\citenamefont
  {Lambert}, \citenamefont {Emary},\ and\ \citenamefont
  {Brandes}}]{Lambert:04}%
  \BibitemOpen
  \bibfield  {author} {\bibinfo {author} {\bibfnamefont {N.}~\bibnamefont
  {Lambert}}, \bibinfo {author} {\bibfnamefont {C.}~\bibnamefont {Emary}}, \
  and\ \bibinfo {author} {\bibfnamefont {T.}~\bibnamefont {Brandes}},\ }\href
  {\doibase 10.1103/PhysRevLett.92.073602} {\bibfield  {journal} {\bibinfo
  {journal} {Phys. Rev. Lett.}\ }\textbf {\bibinfo {volume} {92}},\ \bibinfo
  {pages} {073602} (\bibinfo {year} {2004})}\BibitemShut {NoStop}%
\bibitem [{\citenamefont {Liu}\ \emph {et~al.}(2017)\citenamefont {Liu},
  \citenamefont {Chesi}, \citenamefont {Ying}, \citenamefont {Chen},
  \citenamefont {Luo},\ and\ \citenamefont {Lin}}]{Liu:17}%
  \BibitemOpen
  \bibfield  {author} {\bibinfo {author} {\bibfnamefont {M.}~\bibnamefont
  {Liu}}, \bibinfo {author} {\bibfnamefont {S.}~\bibnamefont {Chesi}}, \bibinfo
  {author} {\bibfnamefont {Z.-J.}\ \bibnamefont {Ying}}, \bibinfo {author}
  {\bibfnamefont {X.}~\bibnamefont {Chen}}, \bibinfo {author} {\bibfnamefont
  {H.-G.}\ \bibnamefont {Luo}}, \ and\ \bibinfo {author} {\bibfnamefont
  {H.-Q.}\ \bibnamefont {Lin}},\ }\href {\doibase
  10.1103/PhysRevLett.119.220601} {\bibfield  {journal} {\bibinfo  {journal}
  {Phys. Rev. Lett.}\ }\textbf {\bibinfo {volume} {119}},\ \bibinfo {pages}
  {220601} (\bibinfo {year} {2017})}\BibitemShut {NoStop}%
\bibitem [{\citenamefont {Shen}\ \emph {et~al.}(2017)\citenamefont {Shen},
  \citenamefont {Yang}, \citenamefont {Wu},\ and\ \citenamefont
  {Zheng}}]{Shen:17}%
  \BibitemOpen
  \bibfield  {author} {\bibinfo {author} {\bibfnamefont {L.-T.}\ \bibnamefont
  {Shen}}, \bibinfo {author} {\bibfnamefont {Z.-B.}\ \bibnamefont {Yang}},
  \bibinfo {author} {\bibfnamefont {H.-Z.}\ \bibnamefont {Wu}}, \ and\ \bibinfo
  {author} {\bibfnamefont {S.-B.}\ \bibnamefont {Zheng}},\ }\href {\doibase
  10.1103/PhysRevA.95.013819} {\bibfield  {journal} {\bibinfo  {journal} {Phys.
  Rev. A}\ }\textbf {\bibinfo {volume} {95}},\ \bibinfo {pages} {013819}
  (\bibinfo {year} {2017})}\BibitemShut {NoStop}%
\bibitem [{\citenamefont {Peng}\ \emph {et~al.}(2019)\citenamefont {Peng},
  \citenamefont {Rico}, \citenamefont {Zhong}, \citenamefont {Solano},\ and\
  \citenamefont {Egusquiza}}]{Peng:19}%
  \BibitemOpen
  \bibfield  {author} {\bibinfo {author} {\bibfnamefont {J.}~\bibnamefont
  {Peng}}, \bibinfo {author} {\bibfnamefont {E.}~\bibnamefont {Rico}}, \bibinfo
  {author} {\bibfnamefont {J.}~\bibnamefont {Zhong}}, \bibinfo {author}
  {\bibfnamefont {E.}~\bibnamefont {Solano}}, \ and\ \bibinfo {author}
  {\bibfnamefont {I.~L.}\ \bibnamefont {Egusquiza}},\ }\href {\doibase
  10.1103/PhysRevA.100.063820} {\bibfield  {journal} {\bibinfo  {journal}
  {Phys. Rev. A}\ }\textbf {\bibinfo {volume} {100}},\ \bibinfo {pages}
  {063820} (\bibinfo {year} {2019})}\BibitemShut {NoStop}%
\bibitem [{\citenamefont {Zhu}\ \emph {et~al.}(2020)\citenamefont {Zhu},
  \citenamefont {Xu}, \citenamefont {Zhang},\ and\ \citenamefont
  {Liu}}]{Zhu:20}%
  \BibitemOpen
  \bibfield  {author} {\bibinfo {author} {\bibfnamefont {H.-J.}\ \bibnamefont
  {Zhu}}, \bibinfo {author} {\bibfnamefont {K.}~\bibnamefont {Xu}}, \bibinfo
  {author} {\bibfnamefont {G.-F.}\ \bibnamefont {Zhang}}, \ and\ \bibinfo
  {author} {\bibfnamefont {W.-M.}\ \bibnamefont {Liu}},\ }\href {\doibase
  10.1103/PhysRevLett.125.050402} {\bibfield  {journal} {\bibinfo  {journal}
  {Phys. Rev. Lett.}\ }\textbf {\bibinfo {volume} {125}},\ \bibinfo {pages}
  {050402} (\bibinfo {year} {2020})}\BibitemShut {NoStop}%
\bibitem [{\citenamefont {Felicetti}\ and\ \citenamefont
  {Le~Boit\'e}(2020)}]{Felicetti:20}%
  \BibitemOpen
  \bibfield  {author} {\bibinfo {author} {\bibfnamefont {S.}~\bibnamefont
  {Felicetti}}\ and\ \bibinfo {author} {\bibfnamefont {A.}~\bibnamefont
  {Le~Boit\'e}},\ }\href {\doibase 10.1103/PhysRevLett.124.040404} {\bibfield
  {journal} {\bibinfo  {journal} {Phys. Rev. Lett.}\ }\textbf {\bibinfo
  {volume} {124}},\ \bibinfo {pages} {040404} (\bibinfo {year}
  {2020})}\BibitemShut {NoStop}%
\bibitem [{\citenamefont {Shen}\ \emph {et~al.}(2021)\citenamefont {Shen},
  \citenamefont {Yang}, \citenamefont {Zhong}, \citenamefont {Yang},\ and\
  \citenamefont {Zheng}}]{Shen:21}%
  \BibitemOpen
  \bibfield  {author} {\bibinfo {author} {\bibfnamefont {L.-T.}\ \bibnamefont
  {Shen}}, \bibinfo {author} {\bibfnamefont {J.-W.}\ \bibnamefont {Yang}},
  \bibinfo {author} {\bibfnamefont {Z.-R.}\ \bibnamefont {Zhong}}, \bibinfo
  {author} {\bibfnamefont {Z.-B.}\ \bibnamefont {Yang}}, \ and\ \bibinfo
  {author} {\bibfnamefont {S.-B.}\ \bibnamefont {Zheng}},\ }\href {\doibase
  10.1103/PhysRevA.104.063703} {\bibfield  {journal} {\bibinfo  {journal}
  {Phys. Rev. A}\ }\textbf {\bibinfo {volume} {104}},\ \bibinfo {pages}
  {063703} (\bibinfo {year} {2021})}\BibitemShut {NoStop}%
\bibitem [{\citenamefont {Dusuel}\ and\ \citenamefont
  {Vidal}(2004)}]{Dusuel:04}%
  \BibitemOpen
  \bibfield  {author} {\bibinfo {author} {\bibfnamefont {S.}~\bibnamefont
  {Dusuel}}\ and\ \bibinfo {author} {\bibfnamefont {J.}~\bibnamefont {Vidal}},\
  }\href {\doibase 10.1103/PhysRevLett.93.237204} {\bibfield  {journal}
  {\bibinfo  {journal} {Phys. Rev. Lett.}\ }\textbf {\bibinfo {volume} {93}},\
  \bibinfo {pages} {237204} (\bibinfo {year} {2004})}\BibitemShut {NoStop}%
\bibitem [{\citenamefont {Vidal}\ and\ \citenamefont
  {Dusuel}(2006)}]{Vidal:06}%
  \BibitemOpen
  \bibfield  {author} {\bibinfo {author} {\bibfnamefont {J.}~\bibnamefont
  {Vidal}}\ and\ \bibinfo {author} {\bibfnamefont {S.}~\bibnamefont {Dusuel}},\
  }\href {http://stacks.iop.org/0295-5075/74/i=5/a=817} {\bibfield  {journal}
  {\bibinfo  {journal} {Europhys. Lett.}\ }\textbf {\bibinfo {volume} {74}},\
  \bibinfo {pages} {817} (\bibinfo {year} {2006})}\BibitemShut {NoStop}%
\bibitem [{\citenamefont {Bastidas}\ \emph {et~al.}(2012)\citenamefont
  {Bastidas}, \citenamefont {Emary}, \citenamefont {Regler},\ and\
  \citenamefont {Brandes}}]{Bastidas:12}%
  \BibitemOpen
  \bibfield  {author} {\bibinfo {author} {\bibfnamefont {V.~M.}\ \bibnamefont
  {Bastidas}}, \bibinfo {author} {\bibfnamefont {C.}~\bibnamefont {Emary}},
  \bibinfo {author} {\bibfnamefont {B.}~\bibnamefont {Regler}}, \ and\ \bibinfo
  {author} {\bibfnamefont {T.}~\bibnamefont {Brandes}},\ }\href {\doibase
  10.1103/PhysRevLett.108.043003} {\bibfield  {journal} {\bibinfo  {journal}
  {Phys. Rev. Lett.}\ }\textbf {\bibinfo {volume} {108}},\ \bibinfo {pages}
  {043003} (\bibinfo {year} {2012})}\BibitemShut {NoStop}%
\bibitem [{\citenamefont {Puebla}\ \emph {et~al.}(2013)\citenamefont {Puebla},
  \citenamefont {Rela\~no},\ and\ \citenamefont {Retamosa}}]{Puebla:13}%
  \BibitemOpen
  \bibfield  {author} {\bibinfo {author} {\bibfnamefont {R.}~\bibnamefont
  {Puebla}}, \bibinfo {author} {\bibfnamefont {A.}~\bibnamefont {Rela\~no}}, \
  and\ \bibinfo {author} {\bibfnamefont {J.}~\bibnamefont {Retamosa}},\ }\href
  {\doibase 10.1103/PhysRevA.87.023819} {\bibfield  {journal} {\bibinfo
  {journal} {Phys. Rev. A}\ }\textbf {\bibinfo {volume} {87}},\ \bibinfo
  {pages} {023819} (\bibinfo {year} {2013})}\BibitemShut {NoStop}%
\bibitem [{\citenamefont {Brandes}(2013)}]{Brandes:13}%
  \BibitemOpen
  \bibfield  {author} {\bibinfo {author} {\bibfnamefont {T.}~\bibnamefont
  {Brandes}},\ }\href {\doibase 10.1103/PhysRevE.88.032133} {\bibfield
  {journal} {\bibinfo  {journal} {Phys. Rev. E}\ }\textbf {\bibinfo {volume}
  {88}},\ \bibinfo {pages} {032133} (\bibinfo {year} {2013})}\BibitemShut
  {NoStop}%
\bibitem [{\citenamefont {Puebla}\ and\ \citenamefont
  {Rela\~no}(2015)}]{Puebla:15}%
  \BibitemOpen
  \bibfield  {author} {\bibinfo {author} {\bibfnamefont {R.}~\bibnamefont
  {Puebla}}\ and\ \bibinfo {author} {\bibfnamefont {A.}~\bibnamefont
  {Rela\~no}},\ }\href {\doibase 10.1103/PhysRevE.92.012101} {\bibfield
  {journal} {\bibinfo  {journal} {Phys. Rev. E}\ }\textbf {\bibinfo {volume}
  {92}},\ \bibinfo {pages} {012101} (\bibinfo {year} {2015})}\BibitemShut
  {NoStop}%
\bibitem [{\citenamefont {Puebla}\ \emph {et~al.}(2017)\citenamefont {Puebla},
  \citenamefont {Hwang}, \citenamefont {Casanova},\ and\ \citenamefont
  {Plenio}}]{Puebla:17}%
  \BibitemOpen
  \bibfield  {author} {\bibinfo {author} {\bibfnamefont {R.}~\bibnamefont
  {Puebla}}, \bibinfo {author} {\bibfnamefont {M.-J.}\ \bibnamefont {Hwang}},
  \bibinfo {author} {\bibfnamefont {J.}~\bibnamefont {Casanova}}, \ and\
  \bibinfo {author} {\bibfnamefont {M.~B.}\ \bibnamefont {Plenio}},\ }\href
  {\doibase 10.1103/PhysRevLett.118.073001} {\bibfield  {journal} {\bibinfo
  {journal} {Phys. Rev. Lett.}\ }\textbf {\bibinfo {volume} {118}},\ \bibinfo
  {pages} {073001} (\bibinfo {year} {2017})}\BibitemShut {NoStop}%
\bibitem [{\citenamefont {Garbe}\ \emph {et~al.}(2017)\citenamefont {Garbe},
  \citenamefont {Egusquiza}, \citenamefont {Solano}, \citenamefont {Ciuti},
  \citenamefont {Coudreau}, \citenamefont {Milman},\ and\ \citenamefont
  {Felicetti}}]{Garbe:17}%
  \BibitemOpen
  \bibfield  {author} {\bibinfo {author} {\bibfnamefont {L.}~\bibnamefont
  {Garbe}}, \bibinfo {author} {\bibfnamefont {I.~L.}\ \bibnamefont
  {Egusquiza}}, \bibinfo {author} {\bibfnamefont {E.}~\bibnamefont {Solano}},
  \bibinfo {author} {\bibfnamefont {C.}~\bibnamefont {Ciuti}}, \bibinfo
  {author} {\bibfnamefont {T.}~\bibnamefont {Coudreau}}, \bibinfo {author}
  {\bibfnamefont {P.}~\bibnamefont {Milman}}, \ and\ \bibinfo {author}
  {\bibfnamefont {S.}~\bibnamefont {Felicetti}},\ }\href {\doibase
  10.1103/PhysRevA.95.053854} {\bibfield  {journal} {\bibinfo  {journal} {Phys.
  Rev. A}\ }\textbf {\bibinfo {volume} {95}},\ \bibinfo {pages} {053854}
  (\bibinfo {year} {2017})}\BibitemShut {NoStop}%
\bibitem [{\citenamefont {Wang}\ \emph {et~al.}(2018)\citenamefont {Wang},
  \citenamefont {You}, \citenamefont {Liu}, \citenamefont {Dong}, \citenamefont
  {Luo}, \citenamefont {Romero},\ and\ \citenamefont {You}}]{Wang:18}%
  \BibitemOpen
  \bibfield  {author} {\bibinfo {author} {\bibfnamefont {Y.}~\bibnamefont
  {Wang}}, \bibinfo {author} {\bibfnamefont {W.-L.}\ \bibnamefont {You}},
  \bibinfo {author} {\bibfnamefont {M.}~\bibnamefont {Liu}}, \bibinfo {author}
  {\bibfnamefont {Y.-L.}\ \bibnamefont {Dong}}, \bibinfo {author}
  {\bibfnamefont {H.-G.}\ \bibnamefont {Luo}}, \bibinfo {author} {\bibfnamefont
  {G.}~\bibnamefont {Romero}}, \ and\ \bibinfo {author} {\bibfnamefont {J.~Q.}\
  \bibnamefont {You}},\ }\href {\doibase 10.1088/1367-2630/aac5b5} {\bibfield
  {journal} {\bibinfo  {journal} {New J. Phys.}\ }\textbf {\bibinfo {volume}
  {20}},\ \bibinfo {pages} {053061} (\bibinfo {year} {2018})}\BibitemShut
  {NoStop}%
\bibitem [{\citenamefont {\ifmmode \check{Z}\else
  \v{Z}\fi{}unkovi\ifmmode~\check{c}\else \v{c}\fi{}}\ \emph
  {et~al.}(2018)\citenamefont {\ifmmode \check{Z}\else
  \v{Z}\fi{}unkovi\ifmmode~\check{c}\else \v{c}\fi{}}, \citenamefont {Heyl},
  \citenamefont {Knap},\ and\ \citenamefont {Silva}}]{Zunkovic:18}%
  \BibitemOpen
  \bibfield  {author} {\bibinfo {author} {\bibfnamefont {B.}~\bibnamefont
  {\ifmmode \check{Z}\else \v{Z}\fi{}unkovi\ifmmode~\check{c}\else
  \v{c}\fi{}}}, \bibinfo {author} {\bibfnamefont {M.}~\bibnamefont {Heyl}},
  \bibinfo {author} {\bibfnamefont {M.}~\bibnamefont {Knap}}, \ and\ \bibinfo
  {author} {\bibfnamefont {A.}~\bibnamefont {Silva}},\ }\href {\doibase
  10.1103/PhysRevLett.120.130601} {\bibfield  {journal} {\bibinfo  {journal}
  {Phys. Rev. Lett.}\ }\textbf {\bibinfo {volume} {120}},\ \bibinfo {pages}
  {130601} (\bibinfo {year} {2018})}\BibitemShut {NoStop}%
\bibitem [{\citenamefont {Hwang}\ \emph {et~al.}(2018)\citenamefont {Hwang},
  \citenamefont {Rabl},\ and\ \citenamefont {Plenio}}]{Hwang:18}%
  \BibitemOpen
  \bibfield  {author} {\bibinfo {author} {\bibfnamefont {M.-J.}\ \bibnamefont
  {Hwang}}, \bibinfo {author} {\bibfnamefont {P.}~\bibnamefont {Rabl}}, \ and\
  \bibinfo {author} {\bibfnamefont {M.~B.}\ \bibnamefont {Plenio}},\ }\href
  {\doibase 10.1103/PhysRevA.97.013825} {\bibfield  {journal} {\bibinfo
  {journal} {Phys. Rev. A}\ }\textbf {\bibinfo {volume} {97}},\ \bibinfo
  {pages} {013825} (\bibinfo {year} {2018})}\BibitemShut {NoStop}%
\bibitem [{\citenamefont {Hwang}\ \emph {et~al.}(2019)\citenamefont {Hwang},
  \citenamefont {Wei}, \citenamefont {Huelga},\ and\ \citenamefont
  {Plenio}}]{Hwang:19}%
  \BibitemOpen
  \bibfield  {author} {\bibinfo {author} {\bibfnamefont {M.-J.}\ \bibnamefont
  {Hwang}}, \bibinfo {author} {\bibfnamefont {B.-B.}\ \bibnamefont {Wei}},
  \bibinfo {author} {\bibfnamefont {S.~F.}\ \bibnamefont {Huelga}}, \ and\
  \bibinfo {author} {\bibfnamefont {M.~B.}\ \bibnamefont {Plenio}},\ }\href
  {https://arxiv.org/abs/1904.09937} {\bibfield  {journal} {\bibinfo  {journal}
  {arXiv:1904.09937}\ } (\bibinfo {year} {2019})}\BibitemShut {NoStop}%
\bibitem [{\citenamefont {Puebla}\ \emph {et~al.}(2020)\citenamefont {Puebla},
  \citenamefont {Smirne}, \citenamefont {Huelga},\ and\ \citenamefont
  {Plenio}}]{Puebla:20}%
  \BibitemOpen
  \bibfield  {author} {\bibinfo {author} {\bibfnamefont {R.}~\bibnamefont
  {Puebla}}, \bibinfo {author} {\bibfnamefont {A.}~\bibnamefont {Smirne}},
  \bibinfo {author} {\bibfnamefont {S.~F.}\ \bibnamefont {Huelga}}, \ and\
  \bibinfo {author} {\bibfnamefont {M.~B.}\ \bibnamefont {Plenio}},\ }\href
  {\doibase 10.1103/PhysRevLett.124.230602} {\bibfield  {journal} {\bibinfo
  {journal} {Phys. Rev. Lett.}\ }\textbf {\bibinfo {volume} {124}},\ \bibinfo
  {pages} {230602} (\bibinfo {year} {2020})}\BibitemShut {NoStop}%
\bibitem [{\citenamefont {Puebla}(2020)}]{Puebla:20b}%
  \BibitemOpen
  \bibfield  {author} {\bibinfo {author} {\bibfnamefont {R.}~\bibnamefont
  {Puebla}},\ }\href {\doibase 10.1103/PhysRevB.102.220302} {\bibfield
  {journal} {\bibinfo  {journal} {Phys. Rev. B}\ }\textbf {\bibinfo {volume}
  {102}},\ \bibinfo {pages} {220302} (\bibinfo {year} {2020})}\BibitemShut
  {NoStop}%
\bibitem [{\citenamefont {Corps}\ and\ \citenamefont
  {Rela\~no}(2021)}]{Corps:21}%
  \BibitemOpen
  \bibfield  {author} {\bibinfo {author} {\bibfnamefont {A.~L.}\ \bibnamefont
  {Corps}}\ and\ \bibinfo {author} {\bibfnamefont {A.}~\bibnamefont
  {Rela\~no}},\ }\href {\doibase 10.1103/PhysRevLett.127.130602} {\bibfield
  {journal} {\bibinfo  {journal} {Phys. Rev. Lett.}\ }\textbf {\bibinfo
  {volume} {127}},\ \bibinfo {pages} {130602} (\bibinfo {year}
  {2021})}\BibitemShut {NoStop}%
\bibitem [{\citenamefont {Zibold}\ \emph {et~al.}(2010)\citenamefont {Zibold},
  \citenamefont {Nicklas}, \citenamefont {Gross},\ and\ \citenamefont
  {Oberthaler}}]{Zibold:10}%
  \BibitemOpen
  \bibfield  {author} {\bibinfo {author} {\bibfnamefont {T.}~\bibnamefont
  {Zibold}}, \bibinfo {author} {\bibfnamefont {E.}~\bibnamefont {Nicklas}},
  \bibinfo {author} {\bibfnamefont {C.}~\bibnamefont {Gross}}, \ and\ \bibinfo
  {author} {\bibfnamefont {M.~K.}\ \bibnamefont {Oberthaler}},\ }\href
  {\doibase 10.1103/PhysRevLett.105.204101} {\bibfield  {journal} {\bibinfo
  {journal} {Phys. Rev. Lett.}\ }\textbf {\bibinfo {volume} {105}},\ \bibinfo
  {pages} {204101} (\bibinfo {year} {2010})}\BibitemShut {NoStop}%
\bibitem [{\citenamefont {Baumann}\ \emph {et~al.}(2010)\citenamefont
  {Baumann}, \citenamefont {Guerlin}, \citenamefont {Brennecke},\ and\
  \citenamefont {Esslinger}}]{Baumann:10}%
  \BibitemOpen
  \bibfield  {author} {\bibinfo {author} {\bibfnamefont {K.}~\bibnamefont
  {Baumann}}, \bibinfo {author} {\bibfnamefont {C.}~\bibnamefont {Guerlin}},
  \bibinfo {author} {\bibfnamefont {F.}~\bibnamefont {Brennecke}}, \ and\
  \bibinfo {author} {\bibfnamefont {T.}~\bibnamefont {Esslinger}},\ }\href
  {\doibase 10.1038/nature09009} {\bibfield  {journal} {\bibinfo  {journal}
  {Nature (London)}\ }\textbf {\bibinfo {volume} {464}},\ \bibinfo {pages}
  {1301} (\bibinfo {year} {2010})}\BibitemShut {NoStop}%
\bibitem [{\citenamefont {Baumann}\ \emph {et~al.}(2011)\citenamefont
  {Baumann}, \citenamefont {Mottl}, \citenamefont {Brennecke},\ and\
  \citenamefont {Esslinger}}]{Baumann:11}%
  \BibitemOpen
  \bibfield  {author} {\bibinfo {author} {\bibfnamefont {K.}~\bibnamefont
  {Baumann}}, \bibinfo {author} {\bibfnamefont {R.}~\bibnamefont {Mottl}},
  \bibinfo {author} {\bibfnamefont {F.}~\bibnamefont {Brennecke}}, \ and\
  \bibinfo {author} {\bibfnamefont {T.}~\bibnamefont {Esslinger}},\ }\href
  {\doibase 10.1103/PhysRevLett.107.140402} {\bibfield  {journal} {\bibinfo
  {journal} {Phys. Rev. Lett.}\ }\textbf {\bibinfo {volume} {107}},\ \bibinfo
  {pages} {140402} (\bibinfo {year} {2011})}\BibitemShut {NoStop}%
\bibitem [{\citenamefont {Mottl}\ \emph {et~al.}(2012)\citenamefont {Mottl},
  \citenamefont {Brennecke}, \citenamefont {Baumann}, \citenamefont {Landig},
  \citenamefont {Donner},\ and\ \citenamefont {Esslinger}}]{Mottl:12}%
  \BibitemOpen
  \bibfield  {author} {\bibinfo {author} {\bibfnamefont {R.}~\bibnamefont
  {Mottl}}, \bibinfo {author} {\bibfnamefont {F.}~\bibnamefont {Brennecke}},
  \bibinfo {author} {\bibfnamefont {K.}~\bibnamefont {Baumann}}, \bibinfo
  {author} {\bibfnamefont {R.}~\bibnamefont {Landig}}, \bibinfo {author}
  {\bibfnamefont {T.}~\bibnamefont {Donner}}, \ and\ \bibinfo {author}
  {\bibfnamefont {T.}~\bibnamefont {Esslinger}},\ }\href {\doibase
  10.1126/science.1220314} {\bibfield  {journal} {\bibinfo  {journal}
  {Science}\ }\textbf {\bibinfo {volume} {336}},\ \bibinfo {pages} {1570}
  (\bibinfo {year} {2012})}\BibitemShut {NoStop}%
\bibitem [{\citenamefont {Jurcevic}\ \emph {et~al.}(2017)\citenamefont
  {Jurcevic}, \citenamefont {Shen}, \citenamefont {Hauke}, \citenamefont
  {Maier}, \citenamefont {Brydges}, \citenamefont {Hempel}, \citenamefont
  {Lanyon}, \citenamefont {Heyl}, \citenamefont {Blatt},\ and\ \citenamefont
  {Roos}}]{Jurcevic:17}%
  \BibitemOpen
  \bibfield  {author} {\bibinfo {author} {\bibfnamefont {P.}~\bibnamefont
  {Jurcevic}}, \bibinfo {author} {\bibfnamefont {H.}~\bibnamefont {Shen}},
  \bibinfo {author} {\bibfnamefont {P.}~\bibnamefont {Hauke}}, \bibinfo
  {author} {\bibfnamefont {C.}~\bibnamefont {Maier}}, \bibinfo {author}
  {\bibfnamefont {T.}~\bibnamefont {Brydges}}, \bibinfo {author} {\bibfnamefont
  {C.}~\bibnamefont {Hempel}}, \bibinfo {author} {\bibfnamefont {B.~P.}\
  \bibnamefont {Lanyon}}, \bibinfo {author} {\bibfnamefont {M.}~\bibnamefont
  {Heyl}}, \bibinfo {author} {\bibfnamefont {R.}~\bibnamefont {Blatt}}, \ and\
  \bibinfo {author} {\bibfnamefont {C.~F.}\ \bibnamefont {Roos}},\ }\href
  {\doibase 10.1103/PhysRevLett.119.080501} {\bibfield  {journal} {\bibinfo
  {journal} {Phys. Rev. Lett.}\ }\textbf {\bibinfo {volume} {119}},\ \bibinfo
  {pages} {080501} (\bibinfo {year} {2017})}\BibitemShut {NoStop}%
\bibitem [{\citenamefont {Cai}\ \emph {et~al.}(2021)\citenamefont {Cai},
  \citenamefont {Liu}, \citenamefont {Zhao}, \citenamefont {Wu}, \citenamefont
  {Mei}, \citenamefont {Jiang}, \citenamefont {He}, \citenamefont {Zhang},
  \citenamefont {Zhou},\ and\ \citenamefont {Duan}}]{Cai:21}%
  \BibitemOpen
  \bibfield  {author} {\bibinfo {author} {\bibfnamefont {M.-L.}\ \bibnamefont
  {Cai}}, \bibinfo {author} {\bibfnamefont {Z.-D.}\ \bibnamefont {Liu}},
  \bibinfo {author} {\bibfnamefont {W.-D.}\ \bibnamefont {Zhao}}, \bibinfo
  {author} {\bibfnamefont {Y.-K.}\ \bibnamefont {Wu}}, \bibinfo {author}
  {\bibfnamefont {Q.-X.}\ \bibnamefont {Mei}}, \bibinfo {author} {\bibfnamefont
  {Y.}~\bibnamefont {Jiang}}, \bibinfo {author} {\bibfnamefont
  {L.}~\bibnamefont {He}}, \bibinfo {author} {\bibfnamefont {X.}~\bibnamefont
  {Zhang}}, \bibinfo {author} {\bibfnamefont {Z.-C.}\ \bibnamefont {Zhou}}, \
  and\ \bibinfo {author} {\bibfnamefont {L.-M.}\ \bibnamefont {Duan}},\ }\href
  {\doibase 10.1038/s41467-021-21425-8} {\bibfield  {journal} {\bibinfo
  {journal} {Nat. Commun.}\ }\textbf {\bibinfo {volume} {12}},\ \bibinfo
  {pages} {1126} (\bibinfo {year} {2021})}\BibitemShut {NoStop}%
\bibitem [{\citenamefont {Pinel}\ \emph {et~al.}(2013)\citenamefont {Pinel},
  \citenamefont {Jian}, \citenamefont {Treps}, \citenamefont {Fabre},\ and\
  \citenamefont {Braun}}]{Pinel2013}%
  \BibitemOpen
  \bibfield  {author} {\bibinfo {author} {\bibfnamefont {O.}~\bibnamefont
  {Pinel}}, \bibinfo {author} {\bibfnamefont {P.}~\bibnamefont {Jian}},
  \bibinfo {author} {\bibfnamefont {N.}~\bibnamefont {Treps}}, \bibinfo
  {author} {\bibfnamefont {C.}~\bibnamefont {Fabre}}, \ and\ \bibinfo {author}
  {\bibfnamefont {D.}~\bibnamefont {Braun}},\ }\href {\doibase
  10.1103/PhysRevA.88.040102} {\bibfield  {journal} {\bibinfo  {journal} {Phys.
  Rev. A}\ }\textbf {\bibinfo {volume} {88}},\ \bibinfo {pages} {040102}
  (\bibinfo {year} {2013})}\BibitemShut {NoStop}%
\bibitem [{\citenamefont {Ferraro}\ \emph {et~al.}(2005)\citenamefont
  {Ferraro}, \citenamefont {Olivares},\ and\ \citenamefont {Paris}}]{Ferraro}%
  \BibitemOpen
  \bibfield  {author} {\bibinfo {author} {\bibfnamefont {A.}~\bibnamefont
  {Ferraro}}, \bibinfo {author} {\bibfnamefont {S.}~\bibnamefont {Olivares}}, \
  and\ \bibinfo {author} {\bibfnamefont {M.}~\bibnamefont {Paris}},\
  }\href@noop {} {\emph {\bibinfo {title} {Gaussian states in quantum
  information}}}\ (\bibinfo  {publisher} {Napoli Bibliopolis},\ \bibinfo {year}
  {2005})\BibitemShut {NoStop}%
\bibitem [{\citenamefont {Polkovnikov}\ and\ \citenamefont
  {Gritsev}(2008)}]{Polkovnikov:08}%
  \BibitemOpen
  \bibfield  {author} {\bibinfo {author} {\bibfnamefont {A.}~\bibnamefont
  {Polkovnikov}}\ and\ \bibinfo {author} {\bibfnamefont {V.}~\bibnamefont
  {Gritsev}},\ }\href {http://dx.doi.org/10.1038/nphys963} {\bibfield
  {journal} {\bibinfo  {journal} {Nat. Phys.}\ }\textbf {\bibinfo {volume}
  {4}},\ \bibinfo {pages} {477} (\bibinfo {year} {2008})}\BibitemShut {NoStop}%
\bibitem [{\citenamefont {Defenu}(2021)}]{Defenu:20}%
  \BibitemOpen
  \bibfield  {author} {\bibinfo {author} {\bibfnamefont {N.}~\bibnamefont
  {Defenu}},\ }\href {https://doi.org/10.1038/s42005-021-00649-6} {\bibfield
  {journal} {\bibinfo  {journal} {Commun. Phys.}\ }\textbf {\bibinfo {volume}
  {4}},\ \bibinfo {pages} {150} (\bibinfo {year} {2021})}\BibitemShut {NoStop}%
\bibitem [{\citenamefont {Abah}\ \emph {et~al.}(2021)\citenamefont {Abah},
  \citenamefont {{De Chiara}}, \citenamefont {Paternostro},\ and\ \citenamefont
  {Puebla}}]{Abah:21}%
  \BibitemOpen
  \bibfield  {author} {\bibinfo {author} {\bibfnamefont {O.}~\bibnamefont
  {Abah}}, \bibinfo {author} {\bibfnamefont {G.}~\bibnamefont {{De Chiara}}},
  \bibinfo {author} {\bibfnamefont {M.}~\bibnamefont {Paternostro}}, \ and\
  \bibinfo {author} {\bibfnamefont {R.}~\bibnamefont {Puebla}},\ }\href
  {https://arxiv.org/abs/2105.00362} {\bibfield  {journal} {\bibinfo  {journal}
  {arXiv:2105.00362}\ } (\bibinfo {year} {2021})}\BibitemShut {NoStop}%
\bibitem [{\citenamefont {Rashid}\ \emph {et~al.}(2016)\citenamefont {Rashid},
  \citenamefont {Tufarelli}, \citenamefont {Bateman}, \citenamefont {Vovrosh},
  \citenamefont {Hempston}, \citenamefont {Kim},\ and\ \citenamefont
  {Ulbricht}}]{Rashid:16}%
  \BibitemOpen
  \bibfield  {author} {\bibinfo {author} {\bibfnamefont {M.}~\bibnamefont
  {Rashid}}, \bibinfo {author} {\bibfnamefont {T.}~\bibnamefont {Tufarelli}},
  \bibinfo {author} {\bibfnamefont {J.}~\bibnamefont {Bateman}}, \bibinfo
  {author} {\bibfnamefont {J.}~\bibnamefont {Vovrosh}}, \bibinfo {author}
  {\bibfnamefont {D.}~\bibnamefont {Hempston}}, \bibinfo {author}
  {\bibfnamefont {M.~S.}\ \bibnamefont {Kim}}, \ and\ \bibinfo {author}
  {\bibfnamefont {H.}~\bibnamefont {Ulbricht}},\ }\href {\doibase
  10.1103/PhysRevLett.117.273601} {\bibfield  {journal} {\bibinfo  {journal}
  {Phys. Rev. Lett.}\ }\textbf {\bibinfo {volume} {117}},\ \bibinfo {pages}
  {273601} (\bibinfo {year} {2016})}\BibitemShut {NoStop}%
\bibitem [{\citenamefont {Breuer}\ and\ \citenamefont
  {Petruccione}(2002)}]{Breuer}%
  \BibitemOpen
  \bibfield  {author} {\bibinfo {author} {\bibfnamefont {H.-P.}\ \bibnamefont
  {Breuer}}\ and\ \bibinfo {author} {\bibfnamefont {F.}~\bibnamefont
  {Petruccione}},\ }\href@noop {} {\emph {\bibinfo {title} {{The theory of open
  quantum systems}}}}\ (\bibinfo  {publisher} {Oxford University Press},\
  \bibinfo {address} {Oxford, UK},\ \bibinfo {year} {2002})\BibitemShut
  {NoStop}%
\end{thebibliography}

%

\onecolumn\newpage
\appendix

\section{Robustness of the squeezing against finite-time cycles and variations in $g_\tau$}\label{app:s}

As commented in the main text, the state becomes squeezed after one cycle, whose squeezing parameter is given by $|s|=\log(3)/2$ (cf. Eq.~\eqref{eq:s}). Here we provide a brief derivation of this expression. For that, we follow~\cite{Defenu:20}. The solution to the dynamics under Eq.~\eqref{eq:H0} can be written in terms of the Ermakov-Milne equation, $\ddot{\xi}(t)+\omega^2(t)\xi(t)=1/(4\xi^3(t))$, where $\omega^2(t)=\omega^2(1-g^2(t))$ is the frequency of the harmonic oscillator at time $t$, and $\xi(t)$ an effective width of the state, whose equilibrium value is $\xi(t)=(2\omega(t))^{-1/2}$. In the $\omega\tau\rightarrow \infty$ limit it is possible to find solutions to the above equation as a combination of Airy functions~\cite{Defenu:20}. Indeed, for $\omega\tau\rightarrow\infty$, one can find the overlap between the ground and the evolved state upon a cycle $g(t)$, which is given by $f(2\tau)=\sin(\pi/3)$ (see Ref.~\cite{Defenu:20} for more details). Since for $\langle x\rangle=\langle p\rangle=0$ the Hamiltonian in Eq.~\eqref{eq:H0} only produces squeezing in the state, and because $|\langle 0|\mathcal{S}(s)|0\rangle|^2=\cosh^{-1}(|s|)$, it follows that $\cosh^{-1}(|s|)=\sin(\pi/3)$ which finally leads to $|s|=\log(3)/2$. 

\begin{figure}[t]
  \centering
  \includegraphics[width=0.8\linewidth]{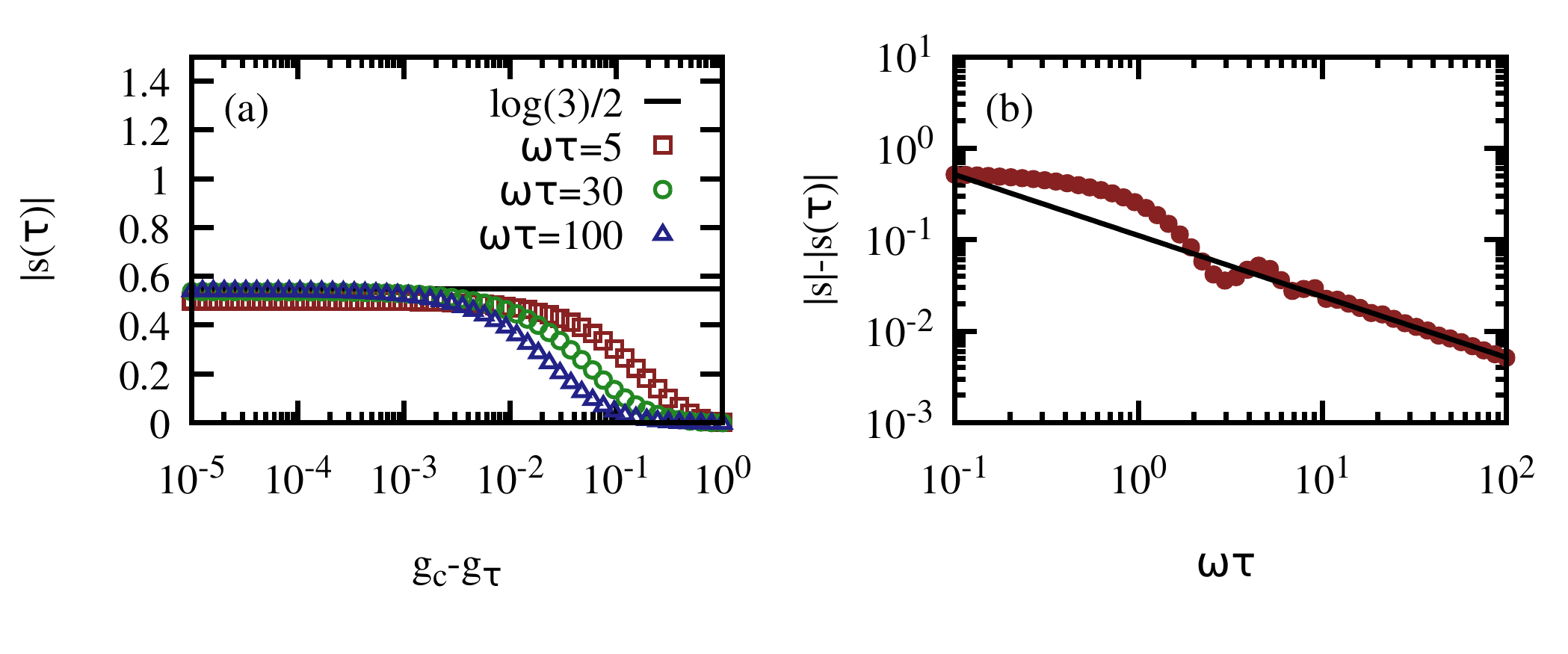}
  \caption{Robustness of the squeezing parameter $|s(\tau)|$ achieved after one cycle under $g(t)$ in a total time $2\tau$, such that $g(0)=0$ and $g(\tau)=g_\tau$, and for an initial vacuum state. Panel (a) shows the squeezing $|s(\tau)|$ as a function of the deviation to the critical point $g_c-g_\tau$. The solid horizontal line corresponds to the expected $|s|=\log(3)/2$, while the points have been obtained numerically for different $\omega\tau$ and $g_\tau$ values. Panel (c) shows the finite-time corrections to the squeezing $|s(\tau)|$ when $g_\tau=1$. The solid line corresponds to a best fit, $|s|-|s(\tau)|\approx (27\omega \tau)^{-2/3}$.}
  \label{figSM:s}
\end{figure}

This result holds to a good degree of approximation for $\omega\tau\gtrsim 1$ and $g_c-g_\tau\ll 1$. Let us denote $|s(\tau)|$ the squeezing parameter after one cycle $g(t)$ (cf. Eq.~\eqref{eq:gt}) with duration $2\tau$ (see App.~\ref{app:dyn} for details on how to obtain $|s|$ from ${\bf R}$). From numerical simulations we find that $|s(\tau)|\approx |s|$ provided $g_c-g_\tau\ll 1$ and $\omega\tau_q\gtrsim 1$. Indeed, Fig.~\ref{figSM:s}(a) shows the robustness of $|s(\tau)|$ when $g_\tau\neq g_c=1$ but $|g_c-g_\tau|\ll 1$. On the one hand, if $g_\tau\ll g_c=1$ the dynamics is not influenced by the critical point, so $|s|\approx 0$. In addition, for $g_\tau\lesssim g_c$ the larger $\tau$, the more sensitive $|s(\tau)|$ becomes to deviations in $g_\tau$. This is due to the finite energy gap at $g<g_c$, that is, for $\tau\gg 1/\Delta(g_\tau)$ the evolution becomes adiabatic and Eq.~\eqref{eq:s} no longer holds. On the other hand, from numerical simulation we find that finite-time corrections when $g_\tau=1$ obey $|s|-|s(\tau)|\approx (27\omega\tau)^{-2/3}$ for $\omega\tau\gtrsim 1$ (cf. Fig.~\ref{figSM:s}(b)). Hence,  by  increasing $\omega\tau$ the resulting squeezing gets closer to the expected $|s|=\log(3)/2$ at the price of loosing robustness against potential deviations from $g_\tau=1$. For $\omega\tau\rightarrow 0$ one trivially obtains $|s(\tau)|\approx 0$. For the numerical results presented in the main text we consider $\omega\tau\sim O(10)$. In particular, for $\omega\tau=8$, the difference between $|s(\tau)|$ and $|s|$ amounts to $\approx 0.03$, and thus, the expected exponential factor $\alpha$, such that $Q_\omega\propto 3^{\alpha m}$, is $\alpha\approx 1.94$ (cf. Fig.~\ref{fig3}(b)) rather than $\alpha=2$ (cf. Eq.~\eqref{eq:Ibscaling}), which would be reached in the limit $\omega\tau\rightarrow\infty$.

\section{Finite-temperature initial state}\label{app:T}
Let us consider an initial thermal state at inverse temperature $\beta=(k_B T)^{-1}$, $\rho_\beta=e^{-\beta H_0}/{\rm Tr}[e^{-\beta H_0}]$ where $H_0=\omega a^\dagger a$ and $N_{\beta}={\rm Tr}[\rho_\beta a^\dagger a]=(e^{\beta\omega}-1)^{-1}$.  As discussed in the main text, the state upon $m$ cycles can acquire a $m$fold squeezing $|s_m|=m|s|$ where $|s|=\log(3)/2$, so that $\rho(2m\tau)=\mathcal{S}(s_m)\rho_\beta \mathcal{S}^\dagger(s_m)$, whose occupation number $N_m={\rm Tr}[\rho(2m\tau)a^\dagger a]$ is given by
\begin{align}
  N_m = \frac{1}{2}\left((2 N_{\beta}+1)\cosh(2|s_m|)-1\right)
  \end{align}
For $N_{\beta}=0$ one recovers the expression in Eq.~\eqref{eq:Nm} when substituting $|s_m|=m\log(3)/2$.  From the bound~\eqref{eq:Ib} and again approximating $2N(t)+1\approx 2N_m+1$ for $t\in[2(m-1)\tau,2m\tau]$ and assuming $|s_m|\gtrsim 1$, we arrive to
\begin{align}
I_\omega^B\approx 16\tau^2 (2N_\beta+1)^2\cosh^2(2|s_m|)\sim \tau^2 3^{T/\tau},
  \end{align}
which is equivalent to Eq.~\eqref{eq:Ibscaling} up to prefactors. The bound captures the exponential scaling. Yet, it becomes loose as it overestimates $I_\omega$, since the $N_\beta$ bosons originally contained in $\rho_\beta$ do not participate actively in the parameter estimation.

\section{Squeezing and quantum Fisher information of a Gaussian state}\label{app:dyn}
A Gaussian state $\rho$ is characterized by a Gaussian Wigner function in the phase space ${\bf X}^\top=(x,p)$, such that $W({\bf X})=P/(2\pi) \ e^{-({\bf X}-\langle {\bf X}\rangle)^\top {\bf R}^{-1}({\bf X}-\langle {\bf X}\rangle)}$, where $\langle {\bf X}^{\top}\rangle=(\langle x\rangle,\langle p\rangle)$ and ${\bf R}$ is the covariance matrix whose matrix elements are $R_{i,j}=\frac{1}{2}\langle X_iX_j+X_jX_i\rangle-\langle X_i\rangle \langle X_j\rangle$. In the previous $P={\rm det}[{\bf R}]^{-1/2}$ denotes the purity of $\rho$~\cite{Pinel2013}. Under the Hamiltonian $H(t)=\omega a^\dagger a-g^2(t)\omega (a+a^\dagger)^2/4$ and the master equation given in Eq.~\eqref{eq:me}, it is straightforward to find the time-dependent Lyapunov equation for ${\bf R}$, which is given in Eq.~\eqref{eq:Lya2}. Recall that we employ $x=a+a^\dagger$ and $p=i(a^\dagger-a)$, so that $N={\rm Tr}[\rho a^\dagger a]=({\rm Tr}[{\bf R}]-2)/4$, while we consider initial states with $\langle {\bf X}^{\top}\rangle=(0,0)$ so that $\langle x\rangle=\langle p\rangle=0\ \forall t$. For a decoherence-free evolution, the covariance matrix ${\bf R}$ fulfills ${\rm det}[{\bf R}]=1$. Moreover, as the evolution produces only squeezing, it can be diagonalized at any time, $\tilde{{\bf R}}={\bf V} {\bf R}{\bf V}^\top$ yielding ${\rm diag}[{\tilde{{\bf R}}}]=(e^{2|s|}, e^{-2|s|})$, while the angle $\theta$ follows from the eigenvectors of ${\bf R}$ in the phase space ${\bf X}$. That is, the eigenvector with eigenvalue $e^{-2|s|}$ is of the form ${\bf v}^\top={\bf x}_s^\top=(\sin(\theta/2),\cos(\theta/2))$.

In all the simulations presented in the main text, the derivatives $\partial_\omega {\bf R}$ and $\partial_\omega P$ have been computed numerically setting $\epsilon/\omega\sim 10^{-8}$ which ensured the convergence of the results. 

The phase $\theta$ upon one cycle can be estimated as follows. First, the phase gained during the evolution introduces a factor $e^{i\beta}$ where $\beta$ reads
\begin{align}
    \beta=-\int_0^{2\tau}dt \ \sqrt{1-g^2(t)}=-2\omega \tau \int_0^1 d\tilde{t} \sqrt{1-g^2(\tilde{t})}
\end{align}
where $g(\tilde{t})$ is the protocol with a rescaled time $\tilde{t}=t/\tau$. That is, the state at time $2\tau$ can be written as $\ket{\psi(2\tau)}=\sum_{n=0}c_ne^{i(n+1)\beta}\ket{n}$. In this manner, we can write $\beta=\omega \tau \nu$ with $\nu=-2\int_0^1d\tilde{t}\ \sqrt{1-g^2(\tilde{t})}$. Then, the phase $\theta$ changes linearly with $\omega\tau$, i.e.
\begin{align}
    \theta=\nu  \omega \tau +\theta_0,
\end{align}
where $\theta_0=\pi/2$ since for $\omega\tau\rightarrow 0$ the evolution squeezes the state in the $x$ direction. For a linear ramp, we find $\nu=-\pi/2$, so that $\theta=-\omega \tau \pi/2+\pi/2$. For $\omega\tau=2n$ with $n=1,2,\ldots$ one obtains $\theta=\pm \pi/2$. For subsequent cycles, the choice $\omega\tau=2n$ leads to the desired phase-matching condition $\theta_{m+1}=\theta_m$.

\end{document}